\documentclass[12pt]{iopart}

\newcommand{\beq}{\begin{equation}}
\newcommand{\eeq}{\end{equation}}
\newcommand{\beqn}{\begin{eqnarray}}
\newcommand{\eeqn}{\end{eqnarray}}

\newcommand{\llabel}[1]{\label{#1}}              % DO NOT show equation label
\newcommand{\labeq}[2]{ \begin{equation} \llabel{#1}{#2}
\end{equation}}
\newcommand{\eqref}[1]{(\ref{#1})}

\def\eadnew#1#2{\address{#2 E-mail: \mailto{#1}}}

\usepackage{graphicx}
\usepackage{epsf}
\usepackage{longtable}
\usepackage{hyperref}
\usepackage{graphics,epsfig,placeins,subfigure,wrapfig}
\usepackage[usenames]{color}
\usepackage{amssymb}
\usepackage{soul}

\begin{document}
\title[Equation of state effects \& one-arm instability in eccentric
  NSNS mergers]{Equation of state effects and one-arm spiral
  instability in hypermassive neutron stars formed in eccentric
  neutron star mergers}

\author{ William E.~East$^{1,2,*}$, Vasileios Paschalidis$^{3,\dag}$,
  Frans Pretorius$^{3,\ddag}$} 
\address{$^1$ Kavli Institute for
  Particle Astrophysics and Cosmology, Stanford University, SLAC
  National Accelerator Laboratory, Menlo Park, California 94025, USA}
\address{$^2$ Perimeter Institute for Theoretical Physics, Waterloo, Ontario N2L 2Y5, Canada}
\address{$^3$ Department of Physics, Princeton University, Princeton,
  NJ 08544, USA} \eadnew{weast@perimeterinstitute.ca}{$^{*}$}
\eadnew{vp16@princeton.edu}{$^{\dag}$}
\eadnew{fpretori@princeton.edu}{$^{\ddag}$}

\begin{abstract}
We continue our investigations of the development and importance of
the one-arm spiral instability in long-lived hypermassive neutron
stars (HMNSs) formed in dynamical capture binary neutron star
mergers. Employing hydrodynamic simulations in full general
relativity, we find that the one-arm instability is generic in that it
can develop in HMNSs within a few tens of
milliseconds after merger for all equations of state in our survey. We
find that mergers with stiffer equations of state tend to produce
HMNSs with stronger $m=2$ azimuthal mode density deformations,
and weaker $m=1$ components, relative to softer equations of state. We
also find that for equations of state that can give rise to
double-core HMNSs, large $m=1$ density modes can
already be present due to asymmetries in the two cores. This results
in the generation of $l=2$, $m=1$ gravitational wave modes even before
the dominance of a one-arm mode that ultimately arises following
merger of the two cores. Our results further suggest that stiffer
equations of state give rise to HMNSs generating lower $m=1$
gravitational wave frequencies. Thus, if gravitational waves from the
one-arm instability are detected, they could in principle constrain
the neutron star equation of state. We estimate that, depending on the
equation of state, the one-arm mode could potentially be detectable by
second generation gravitational wave detectors at $\sim 10$ Mpc and by
third generation ones at $\sim 100$ Mpc. Finally, we provide estimates
of the properties of dynamical ejecta, as well as the accompanying
kilonovae signatures.

\end{abstract}

% Let's try and keep it on one page, with the Title, Authors, & Abstract:
% Optional additional PACS:
%04.30.-w, 04.70.-s, 04.40.Nr, 52.27.Ny,

\maketitle

\section{Introduction}

The LIGO and Virgo collaboration recently ushered in the era of gravitational
wave (GW) astronomy with the observation of two high confidence signals that are both 
consistent with the inspiral and merger of binary black hole
systems \cite{LIGO_first_direct_GW,LIGOseconddirect}. This important
milestone provides spectacular confirmation of the predictions of
general relativity in the dynamical strong-field regime,
the cleanest evidence yet for the existence of black holes, and gives
important clues to the evolution of massive stars from inferred masses
and limits on spins of the constituent black holes involved in the mergers.
Over the next few years, it is anticipated that GWs will be observed not only from
additional black hole binary mergers, but also from neutron
star--neutron star (NSNS) and black hole--neutron star (BHNS)
binary mergers.
 
Coalescing NSNSs and BHNSs are not only important sources of GWs, but may
also be the progenitor systems that power short gamma-ray
bursts~\cite{EiLiPiSc,NaPaPi,MoHeIsMa,Berger2014,PaschalidisJet2015}. In
addition, these systems could generate other detectable
electromagnetic (EM) signals, prior
to~\cite{Hansen:2000am,McWilliams:2011zi,Paschalidis:2013jsa,
  PalenzuelaLehner2013,2014PhRvD..90d4007P,Metzger2016MNRAS.461.4435M} and
following~\cite{MetzgerBerger2012,2015MNRAS.446.1115M} merger.
Combining GW and EM signals observed from such events could provide
further precision tests of relativistic gravity, constrain the NS equation of
state (EOS), and may help explain where the r-process
elements in the Universe come from~\cite{Rosswog:1998gc}. However, the
interpretation of such ``multimessenger'' signals from compact
binaries will be limited by our theoretical understanding of the
processes involved. Gaining the requisite knowledge requires simulations in full
general relativity (GR) because of the crucial role relativistic
gravitation plays in such systems. 

There have been multiple studies of compact binary mergers in full GR.
For NSNSs, see~\cite{faber_review,Baiotti_Rezzolla_review2016} for
reviews and~\cite{Paschalidis2012,gold,East2012NSNS,Bernuzzi:2013rza,
  Neilsen2014,2014PhRvD..90d1502K,2015arXiv150707100D,Dionysopoulou2015,Dietrich2015,2015arXiv151206397B,
  2015PhRvD..91f4059S,2015PhRvD..92l4034K,Palenzuela2015,Kastaun2015,2015arXiv151206397B,2016PhRvD..93d4019F,2016arXiv160400782H,2016PhRvD..93f4082H,RLPSNSNSjet2016,2016arXiv160403445E,2016MNRAS.tmp..894R,Dietrich2016arXiv160706636D,2016arXiv160301918S}
for some more recent results. These earlier works have advanced our
knowledge of EOS effects, neutrino transport, ejecta properties, jet
outflows and magnetospheric phenomena. In addition, several studies
(see
e.g.,~\cite{Stergioulas:2011gd,Bauswein2015a,Bauswein2015b,Bauswein2015c,Takami2014,Takami2015})
have investigated how GWs generated by oscillations of a hypermassive
neutron star (HMNS) formed post-merger can be used to infer the
nuclear EOS. These studies have focused primarily on the
$l=2$, $m=2$ GW mode, which in terms of energy flux is the dominant
component of the GW emission. However, the frequency of this mode is
of order 2-3 kHz, outside the range where the advanced gravitational
wave detectors have the most sensitivity.

In a recent work of ours~\cite{PEPS2015}, we discovered that a
so-called one-arm spiral instability can develop in HMNSs formed in
high-eccentricity dynamical capture binary neutron star
mergers. Heuristically speaking, once saturated the one-arm
instability manifests as a dense core offset from the center of mass
about which it rotates at roughly constant frequency, and the $m=1$
azimuthal deformation of the rest-mass density distribution dominates
over all other non-zero $m$ modes. For more details on the history and
features of the one arm spiral instability, and a discussion of
dynamical capture versus field binaries, we refer the reader to
~\cite{PEPS2015,EPPS2016} and the references therein.  This $m=1$
deformation of the rest-mass density generates $l=2$, $m=1$ GW
modes. These $(2,\ 1)$ GW modes from the instability are not initially
as strong as the $(2,\ 2)$ modes, but they have almost constant
frequency and amplitude compared to the decaying $(2,\ 2)$ modes.  Moreover,
the $m=1$ nature of these modes means that their GW frequency is half that of
the corresponding $m=2$ modes, lying in a band where ground-based gravitational
wave detectors are more sensitive.  Hence if the HMNS and conditions driving
the instability survive for long enough, the $(2,\ 1)$ modes could eventually
dominate the GW signal-to-noise. 

In~\cite{EPPS2016} we considered a range of NS spins and impact
parameters, all corresponding to initially marginally unbound
($e\approx 1$) equal mass ($1.35M_\odot$ each) systems, and chose a
single EOS. For this range of parameters we found that the one-arm instability develops when
the total angular momentum at merger divided by the total mass squared is
$J/M^2 \sim 0.9-1.0$. This range is relevant for
quasi-circular NSNS binaries, and so we conjectured that the one
arm instability should also manifest in qausicircular NSNS binaries 
if a HMNS forms post-merger. 
This was recently confirmed first by \cite{2016arXiv160305726R}, and
subsequently by \cite{2016arXiv160502369L} who studied quasi-circular
mergers of equal-mass and unequal-mass irrotational NSNSs,
respectively.

In this work, we continue our investigation of the relevance of the
one-arm instability in HMNS remnants following eccentric NS mergers by
expanding the parameter space of initial conditions and NS properties.
In particular, we consider
different equations of state, as well as unequal mass stars and
different total masses for the binary.  Our survey indicates that the
one-arm spiral instability is generic in that it can be excited for
all of the equations of state in our survey which form sufficiently
long-lived HMNSs (i.e. which do not promptly collapse to black
holes)\footnote{Though note that here we focus on impact parameters
  and initial NS spins that satisfy the criterion $J/M^2 \sim 0.9-1.0$
  prior to merger, identified in~\cite{EPPS2016} as seeming to
  indicate when the remnant would be subject to the instability.
  Presumably here for the different EOSs and mass ratios similar
  criteria hold.}.  At this point it is important to clarify that
while the term ``instability'' would normally imply a mode whose
amplitude grows from some initial small seed, throughout we will use
the term ``one-arm spiral instability'' to also imply the dominance of
a ``one-arm mode'' (a nearly constant amplitude $m=1$ density mode)
in the final HMNS, even when the initial $m=1$ ``seed''
is large,
%FP: This sentence was really long, so trimmed the following as I think "large" is enough:
%and potentially as large as the saturation amplitude of the
%one-arm mode in the final HMNS, 
for e.g. when the binary mass
ratio deviates significantly from unity.

%For stiffer equations of state we find
%that the $m=1$ azimuthal deformation of the HMNS
%dominates over all other non-zero $m$ modes on a much longer
%timescale. 
We find that mergers with stiffer EOSs tend to produce HMNSs with
stronger $m=2$ density deformations, and weaker $m=1$ components
relative to softer EOSs.  However, even for the stiffest EOS
considered here, the HMNS can be dominated by a one-arm mode within a
few tens of ms after merger. We also show that the EOS dependency of
the size and oscillations of the resulting HMNS is imprinted in the
frequency of the resulting GWs, which are sharply peaked in frequency
space.  In addition, we discover that the one-arm mode can dominate
not only for equations of state that lead to toroidal and ellipsoidal
HMNSs, but also for equations of state in which a double-core HMNS
forms following merger.  In this case, the one-arm mode dominates only
after the two cores merge into one. Prior to merger of the two cores,
the $m=2$ density mode dominates in the cases considered
here. However, small $m=1$ asymmetries at merger give rise to
asymmetric cores in the double-core remnant which, like the one-arm
mode, may act as quasi-stationary sources of $l=2$, $m=1$ GW modes. 

As a side product we also estimate the properties of dynamically ejected
matter, and find that for a given periapse distance unequal mass ratios
eject more matter and the associated kilonovae signatures are
brighter.

The remainder of the paper is organized as follows. In Section~\ref{Methods} we
describe the equations of state we survey, the initial data, and the methods we
adopt for evolving the Einstein and general relativistic hydrodynamic
equations; in Sec.~\ref{Results} we present the results from our simulations;
and in Sec.~\ref{Conclusions} we conclude with a summary of our main findings
and description of future work.  Unless otherwise specified, below we adopt
geometrized units where $G=c=1$.

\section{Methods}
\label{Methods}

\subsection{Initial conditions}
As in~\cite{idsolve_paper,PEPS2015}, we construct
constraint-satisfying initial data for our evolutions starting from
isolated rigidly-rotating equilibrium NS solutions that are generated
with the code of~\cite{1994ApJ...424..823C,1994ApJ...422..227C}. The
free-data for the metric and matter fields are then determined by
superposing two boosted isolated NS solutions, with the velocities and
positions of a marginally unbound Newtonian orbit at a separation of
$d=50M$ ($\sim200$ km; where $M$ is the total ADM mass), and then
solving the constraints. Here we consider the piece-wise polytropic
equations of state labelled ``B", ``HB", ``H", ``2H" in~\cite{read},
which yield maximum masses for nonspinning neutron stars --- the
Tolman-Oppenheimer-Volkov (TOV) limit --- of 2.06, 2.12, 2.25, and 2.83
$M_\odot$, respectively. Of these equations of state, ``B'' is the
softest, while ``2H'' is the stiffest one, where a stiffer equation of
state yields a non-rotating neutron star of larger areal radius for
the same gravitational mass. In addition, we consider a $\Gamma=3$
polytropic EOS. For the $\Gamma=3$ EOS specifically, we choose two
different compaction stars, which when assigned the same gravitational
mass ($1.35M_\odot$) can be viewed as stars constructed with different
equations of state corresponding to the same polytropic index, but
different values for the polytropic constant. In particular, the
values we choose are $k_1=349981\ \rm km^4$, yielding a maximum TOV
mass of $2.06M_\odot$, and $k_2=460254\ \rm km^4$, yielding a maximum
TOV mass of $2.21M_\odot$. Throughout we will refer to these as
equations of state $k_1$ and $k_2$ respectively.  To account for shock
heating following merger, we also add a thermal component to the
pressure: $P_{\rm th}=(\Gamma_{\rm th}-1)\epsilon_{\rm th}\rho_0$
where $\epsilon_{\rm th}$ is the thermal part of the internal specific
energy $\epsilon$ and $\rho_0$ is the rest mass density. We choose
$\Gamma_{\rm th}=1.5$ for all cases, except for the $\Gamma=3$
polytropic stars, which we evolve with a $\Gamma$-law EOS
$P=(\Gamma-1)\epsilon\rho_0$ with $\Gamma=3.0$\footnote{Though
  technically such an EOS can give superluminal sound speeds at
  sufficiently large values of $\epsilon$, we have checked that this
  does not occur for the cases considered here.}.

In Table~\ref{ns_table} we list several properties of the NS models we
consider in this work. The spins of the stars we consider correspond
to rotation periods of $\sim 7-15$ ms which are the most likely rotation
periods of observed pulsars in globular clusters~\cite{PEPS2015}. 
The ratio of kinetic to gravitational potential energy is $< 0.11$ for
the rotating equilibria considered here, and thus all of these models
are stable against the development both of the dynamical and the
secular bar-mode
instability~\cite{DynamicalBarmodeOrig,StergioulasReview}. 

For the binary simulations considered here, we restrict ourselves to
cases where the NS spin is aligned with the orbital angular momentum
of the system.

%%%%%%%%%%%%%%%%%%%%%%%%%%%%%%%%%%%%%%%%%%%%%%%%%%%%%%%%%%%%%%%%%%%%%%%%%%%%
\begin{table}[t]
\centering
\caption{\label{ns_table} Properties of isolated NS models considered
  in this work. Listed are the equation of state, dimensionless NS
  spin $a_{\rm NS}$, spin period $P_{\rm s}$ in ms, rest mass $M_0$ in
  $M_\odot$, ADM mass $M_{\rm NS}$ in $M_\odot$, circumferential
  equatorial radius $R_{\rm NS}$ in km, compaction $C=M_{\rm
    NS}/R_{\rm NS}$, and ratio of kinetic $T$ to potential $|W|$
  energy.}
\begin{tabular}{cccccccc}
\hline\hline
EOS &
$a_{\rm NS}$ &
$P_{\rm s}({\rm ms})$ &
$M_0(M_\odot)$ & 
$M_{\rm NS}(M_\odot)$ & 
$R_{\rm NS}({\rm km})$ &
$C$ &
$\frac{T}{|W|}\times 100$ 
\\
\hline
   B  & 0.075 &  6.35 & 1.50 & 1.35 & 10.97 & 0.18 &  0.15 \\
   B  & 0.05  &  9.50 & 1.50 & 1.35 & 10.95 & 0.18 &  0.07 \\
   HB & 0.05  & 10.85 & 1.41 & 1.28 & 11.60 & 0.16 &  0.06 \\
   HB & 0.05  & 10.10 & 1.58 & 1.42 & 11.59 & 0.18 &  0.07 \\
   H  & 0.075 &  7.71 & 1.48 & 1.35 & 12.28 & 0.16 &  0.14 \\
   H  & 0.05  & 11.54 & 1.48 & 1.35 & 12.26 & 0.16 &  0.06 \\
   2H & 0.075 & 11.28 & 1.45 & 1.35 & 15.23 & 0.13 &  0.13 \\
   2H & 0.05  & 16.88 & 1.46 & 1.35 & 15.21 & 0.13 &  0.06 \\
   2H & 0.05 & 17.46 & 1.37 & 1.28 & 15.17 & 0.12 &  0.05 \\
   2H & 0.05 & 16.28 & 1.54 & 1.42 & 15.25 & 0.14 &  0.06 \\
   2H & 0.075 &  9.72 & 1.87 & 1.70 & 15.37 & 0.16 &  0.14 \\
   2H & 0.05  & 14.54 & 1.87 & 1.70 & 15.35 & 0.16 &  0.06 \\
$k_1$ & 0.075 &  6.68 & 1.51 & 1.35 & 10.88 & 0.18 &  0.15 \\
$k_1$ & 0.05  &  9.98 & 1.51 & 1.35 & 10.87 & 0.18 &  0.07 \\
$k_2$ & 0.075 &  7.49 & 1.50 & 1.35 & 11.62 & 0.17 &  0.14 \\
$k_2$ & 0.05  & 11.21 & 1.50 & 1.35 & 11.60 & 0.17 &  0.06 \\
\hline\hline 
\end{tabular}
\end{table}
%%%%%%%%%%%%%%%%%%%%%%%%%%%%%%%%%%%%%%%%%%%%%%%%%%%%%%%%%%%%%%%%%%%%%%%%%%%%

\subsection{Evolution techniques and diagnostics}

We evolve the general relativistic hydrodynamic equations with the
code of~\cite{code_paper}, which solves the Einstein field equations
in the generalized-harmonic formulation with fourth-order accurate
finite differences, and the hydrodynamic equations in conservative
form using finite volume techniques as detailed
in~\cite{bhns_astro_paper}.

In the analysis below, and in particular for studying the evolution of
the one arm instability in a HMNS that forms post-merger, we will use
the complex azimuthal mode decomposition of the conserved rest-mass
density integrated throughout the star
\labeq{Cm}{
C_m = \int \rho_0u^0\sqrt{-g} e^{im\phi}d^3x,
}
where $u^\mu$ is the fluid 4-velocity, $g$ the determinant of the
spacetime metric, and $\rho_0$ the rest-mass density. We also track the
coefficients $C_{lm}(t,r)$ of the expansion of the Newman-Penrose scalar $\Psi_4$ in 
spin-weighted spherical harmonics~\cite{BSNRbook} defined as
\labeq{Clm}{
C_{lm}(t,r)=\int d\Omega\ _{-2}Y_{lm}^*(\theta,\phi) \Psi_4(t,r,\theta,\phi),
}
where $d\Omega$ stands for the differential solid angle,
$_{s}Y_{lm}^*$ the complex conjugate of the spin-weighted spherical
harmonics $_{s}Y_{lm}$, with $s=-2$ here.

\section{Results}
\label{Results}

\subsection{Equation of state study in equal mass mergers}

\begin{figure}
\raggedleft
\includegraphics[trim =0.25cm 0.2cm 2.5cm  0.2cm,clip=true,width=0.49\textwidth]{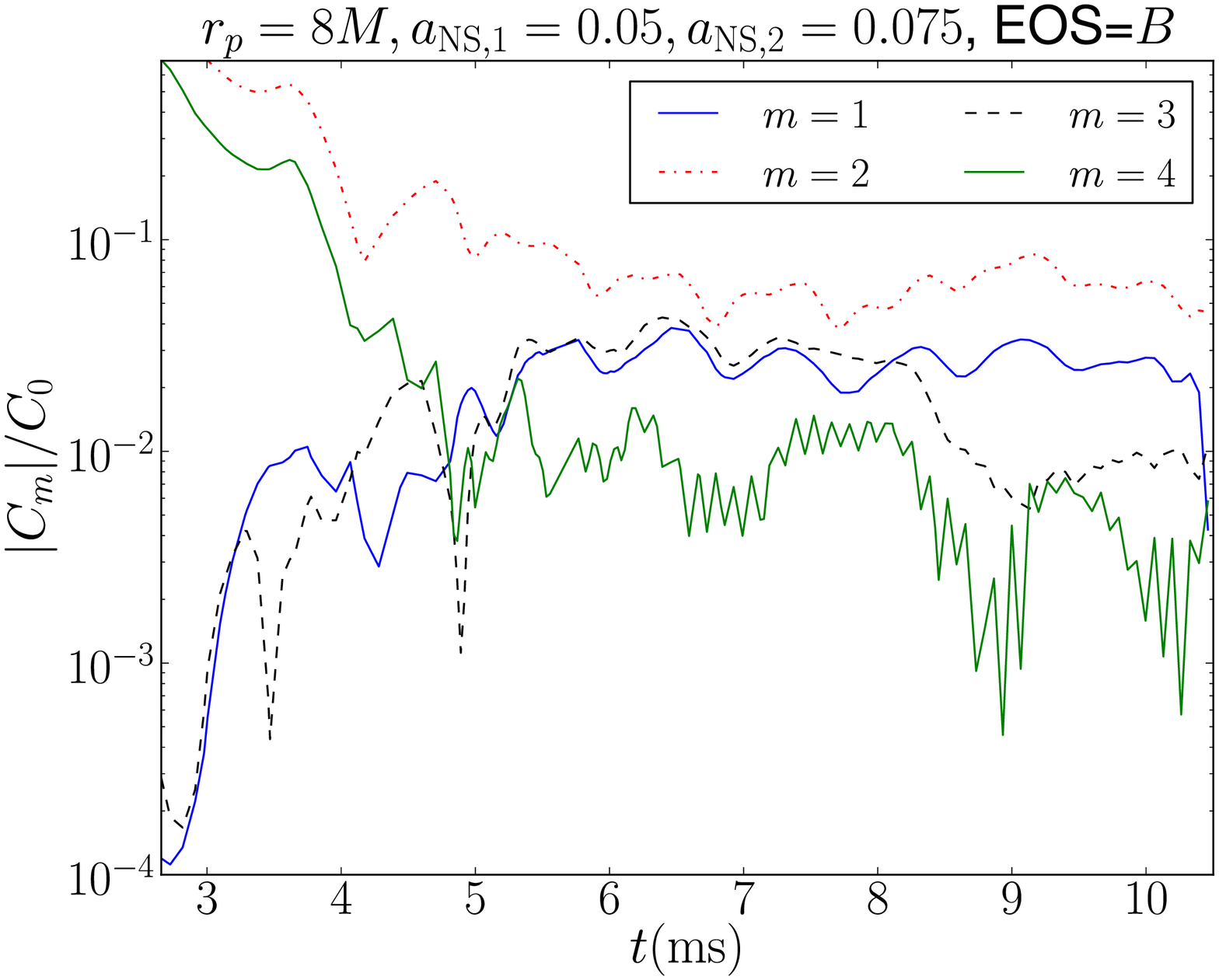}
\includegraphics[trim =0.25cm 0.2cm 2.5cm  0.2cm,clip=true,width=0.49\textwidth]{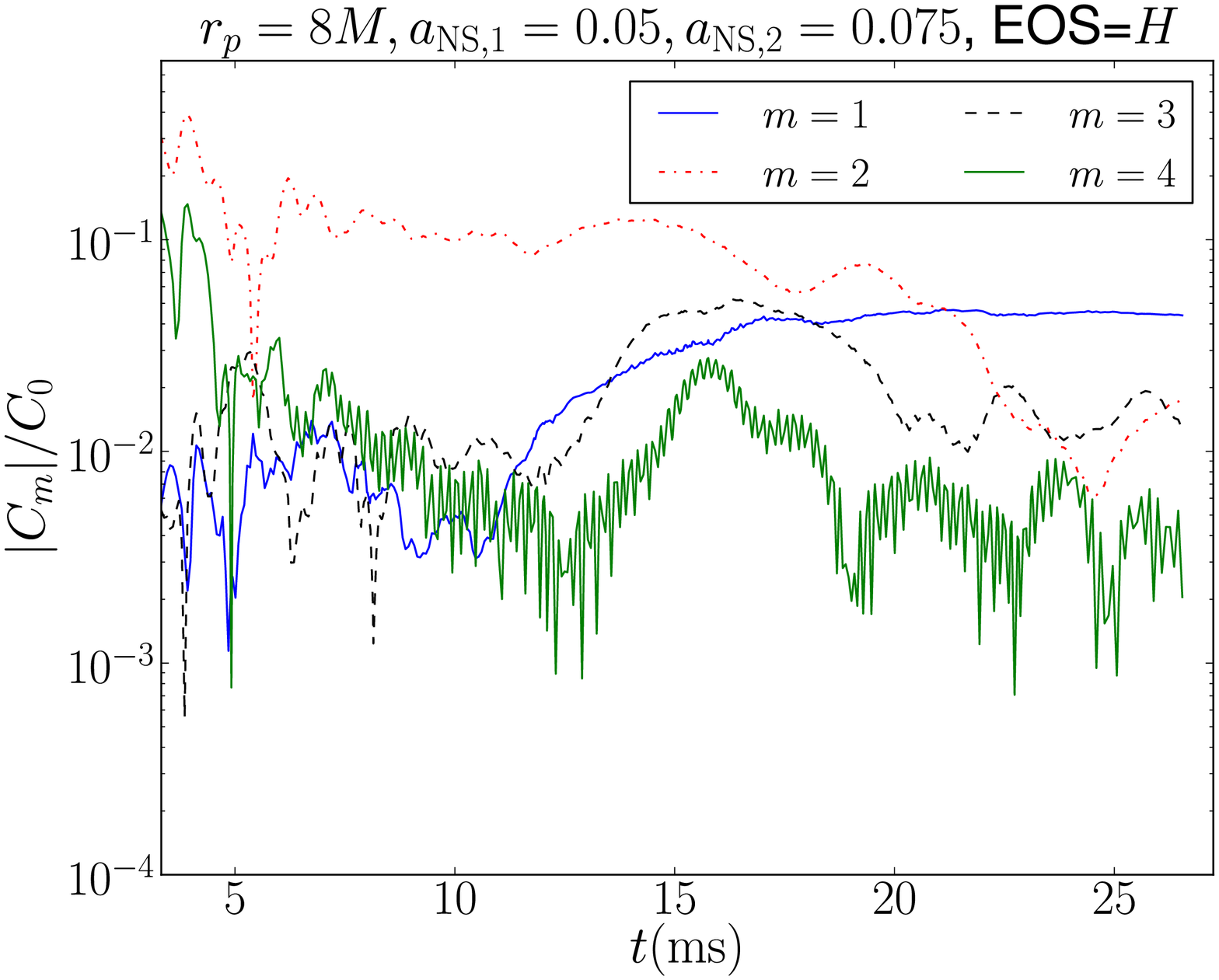}
\includegraphics[trim =0.25cm 0.2cm 2.5cm 0.2cm,clip=true,width=0.49\textwidth]{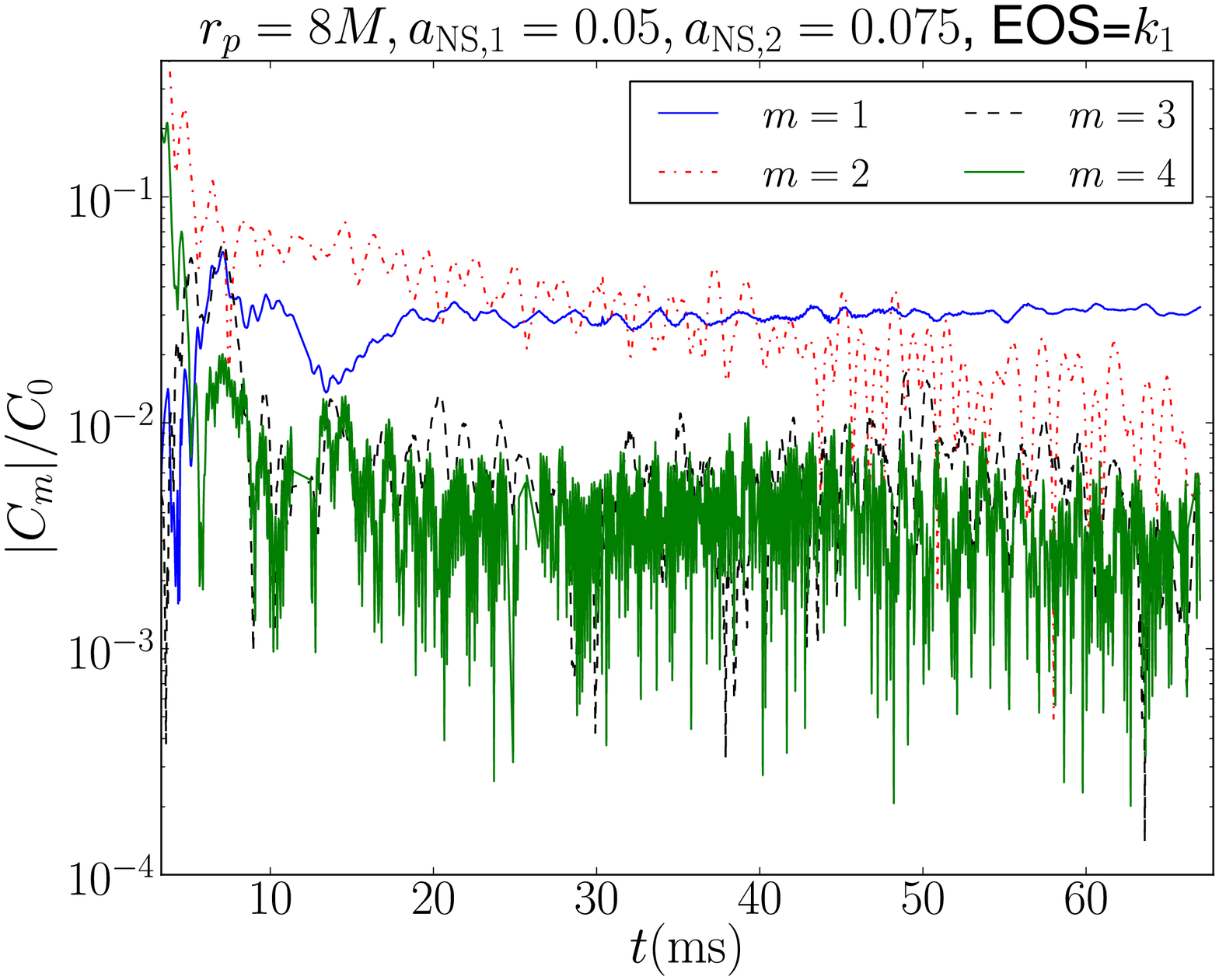}
\includegraphics[trim =0.25cm 0.2cm 2.5cm  0.2cm,clip=true,width=0.49\textwidth]{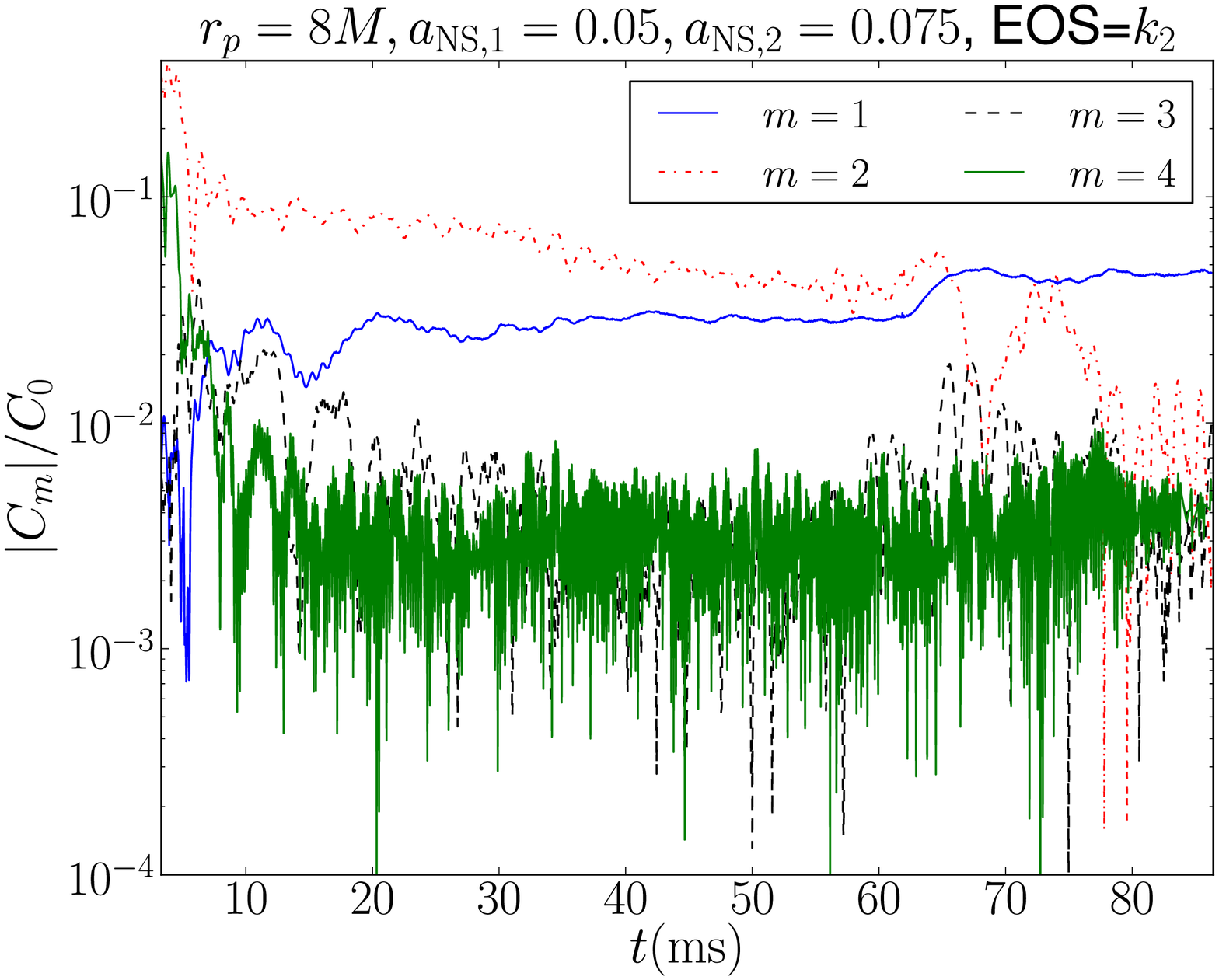}
\caption{Amplitude of $C_m$ normalized to $C_0$ for various $q=1$
  cases at $r_p=8M$. 
  (Note each panel covers a different time range).
  }\label{modesEOSstudy}
\end{figure}

\begin{figure}
\raggedleft
\includegraphics[trim =0.25cm 0.2cm 0.5cm  0.2cm,clip=true,width=0.49\textwidth]{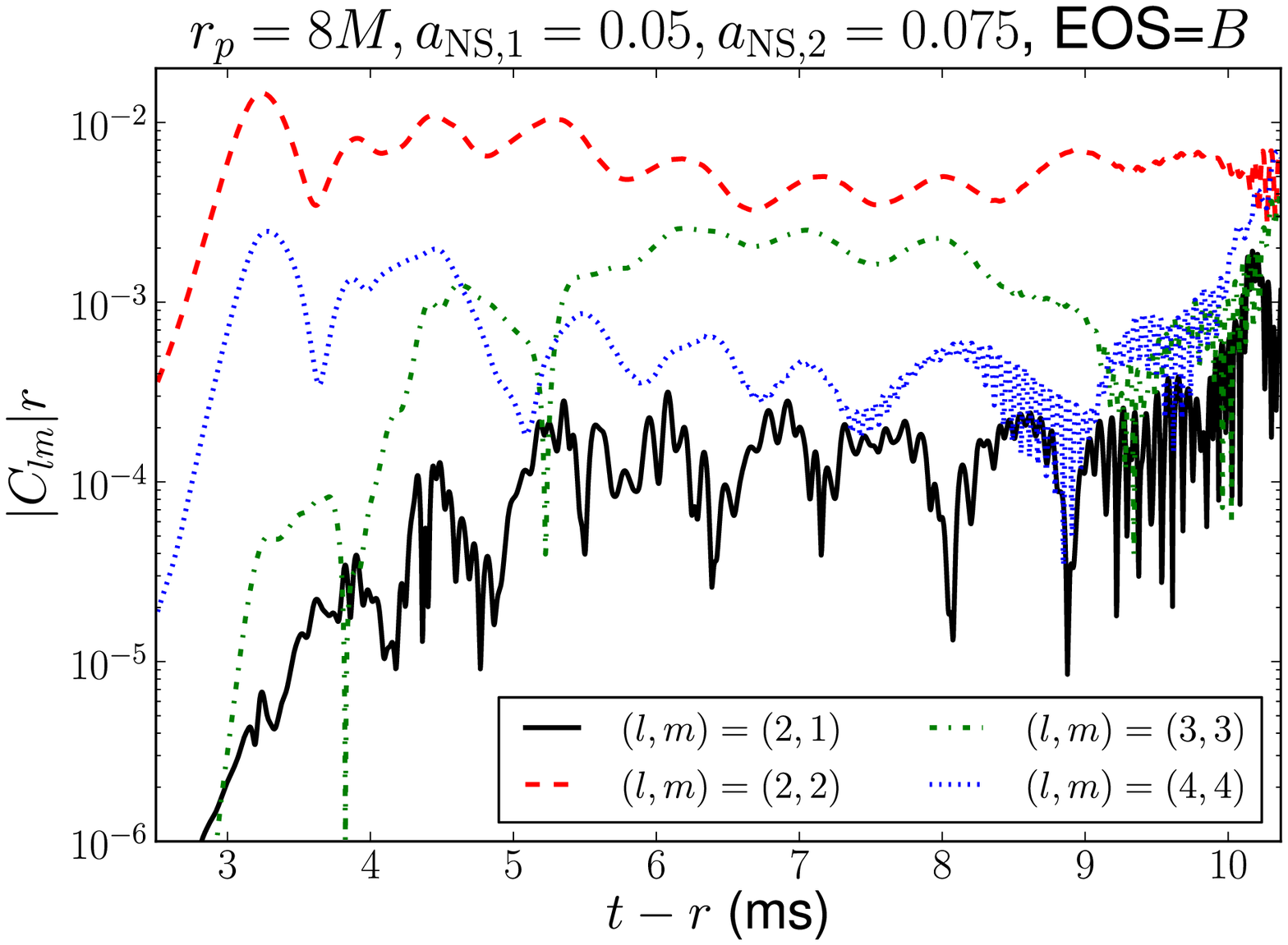}
\includegraphics[trim =0.25cm 0.2cm 0.5cm  0.2cm,clip=true,width=0.49\textwidth]{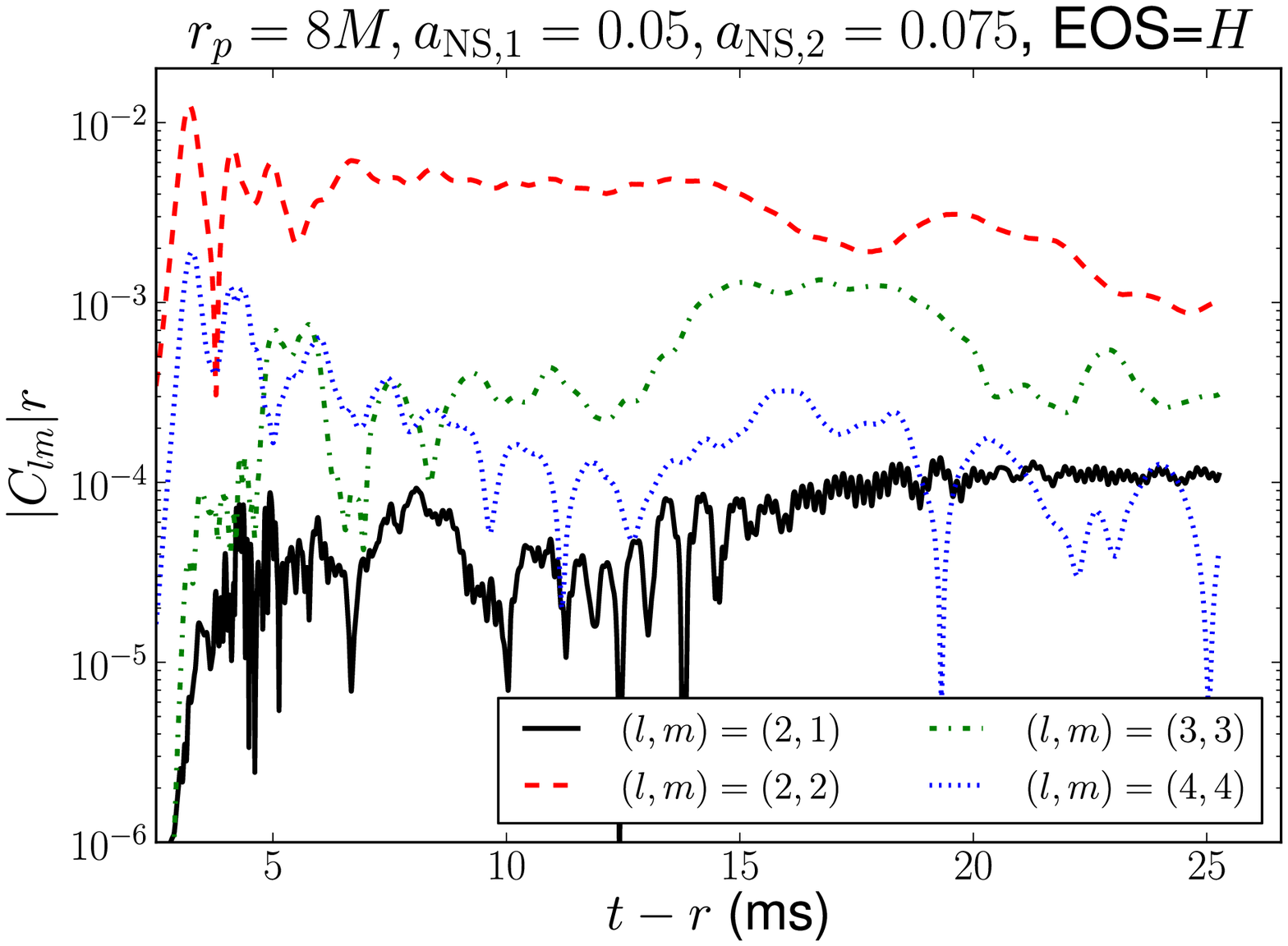}
\includegraphics[trim =0.25cm 0.2cm 0.5cm  0.2cm,clip=true,width=0.49\textwidth]{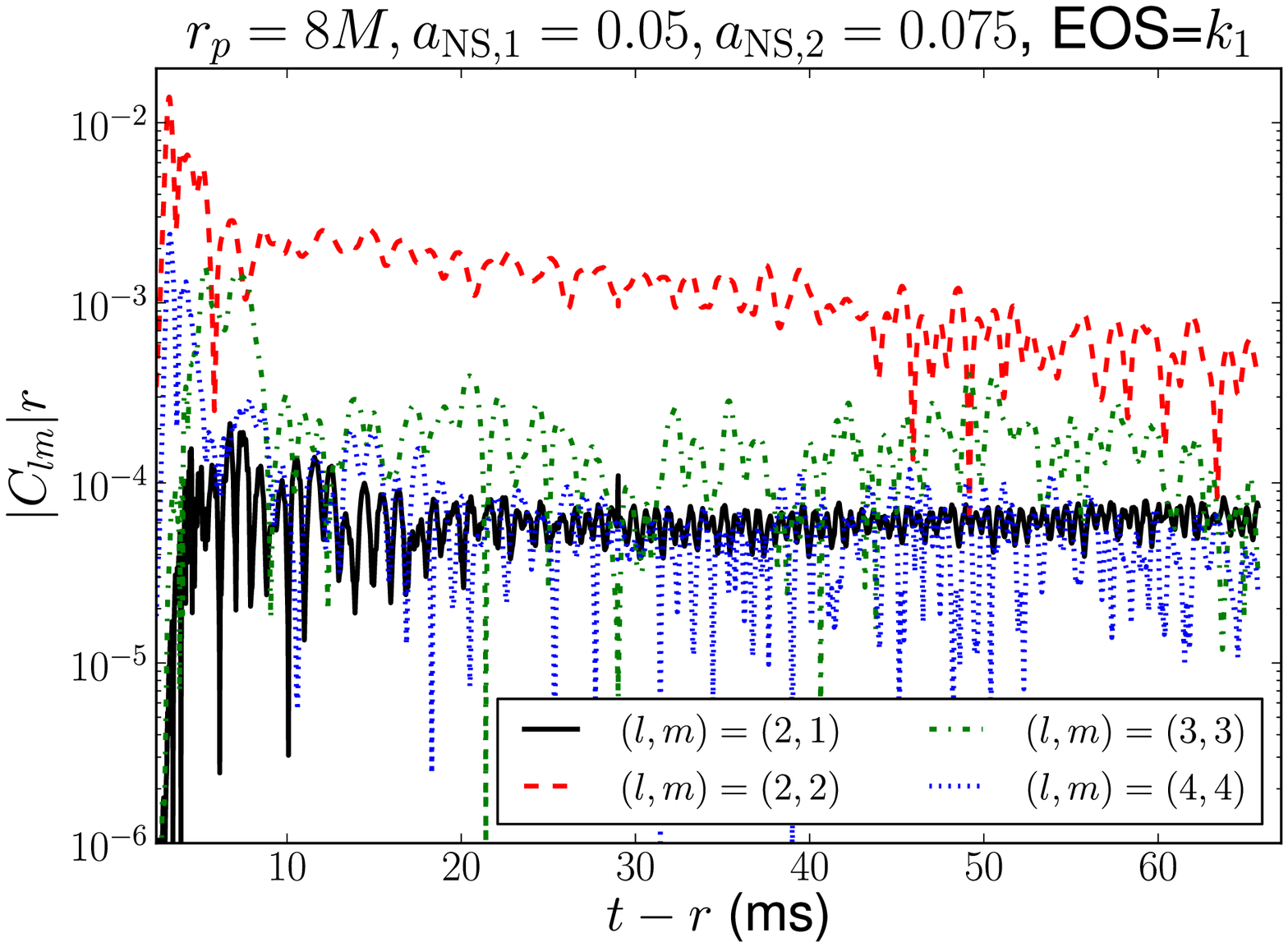}
\includegraphics[trim =0.25cm 0.2cm 0.5cm  0.2cm,clip=true,width=0.49\textwidth]{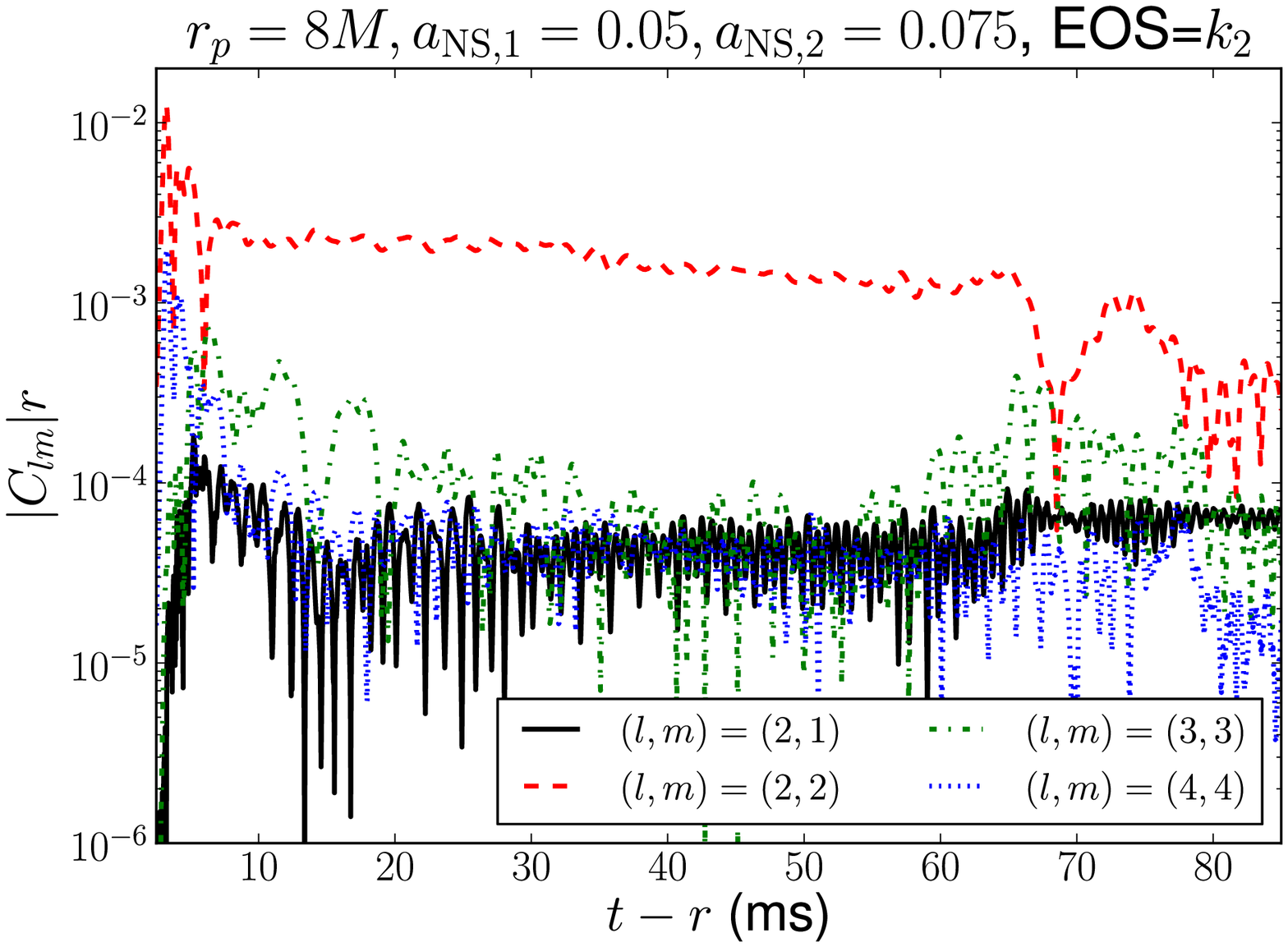}
\caption{Amplitude of several spherical harmonic components of the GW
  signal, normalized by $M$, following merger. Each panel is from a case 
  as shown in the corresponding panel of Fig.~\ref{modesEOSstudy}.
  }\label{gwclmEOSstudy}
\end{figure}

To begin with, we consider initial configurations in which each NS has
a gravitational mass in isolation of $1.35M_\odot$, and initial
orbital parameters corresponding to a (Newtonian orbit) periapse
distance $r_p=8M$, and vary the EOS.  We introduce a small asymmetry
by letting one of the stars have a dimensionless spin of $a_{\rm
  NS,1}=0.05$, and the other $a_{\rm NS,2}=0.075$.  For almost all the
cases considered here, we find that merger produces a long-lived
hypermassive star that lasts the lifetime of the simulation, and
furthermore that the $m=1$ instability is excited and eventually
saturates in the same fashion as described in the introduction. The
merger with the B EOS (the softest considered here) is the one
exception that collapses to a BH during the time of the simulation,
after $\sim 10$ ms.  In Fig.~\ref{modesEOSstudy} we plot the amplitude
of the azimuthal mode decomposition of the HMNS density versus time
for some example cases, where it is clearly shown that, with the
exception of the B EOS case, for sufficiently long evolutions the
$m=1$ azimuthal modes dominate over all other non-zero modes. We plot
the corresponding $C_{lm}$'s of the GWs in Fig.~\ref{gwclmEOSstudy}.
Even when the $m=1$ mode dominates the density decomposition, it is
still the $l=2$, $m=2$ component of the oscillating HMNS that
initially radiates the strongest GWs. This is at least in part
responsible for the decay of the $l=2$, $m=2$ mode with time. In
contrast, in all cases where a clear one-arm mode dominates in the
resulting HMNS, we see no evidence that the $l=2$, $m=1$ GW mode
decays; this is to be expected as long as conditions persist where the
corresponding density mode of the NS is unstable.  By the end of
several of the simulations (including the unequal mass cases
illustrated in Fig.~\ref{gwclmEOSstudy_q09}) we find the $C_{22}$
component is only a factor of 1-4 times that of $C_{21}$.  This
implies that the $l=2$, $m=1$ component of the GW strain $h_{21}$ can
become comparable to or larger than the $l=2$, $m=2$ component
$h_{22}$; similarly for the energy $E_{21}$ vs $E_{22}$ carried by the
corresponding GW modes. This is because for these near monochromatic
GW sources $h_{21}/h_{22} \propto (\Omega_{22}/\Omega_{21})^2
(C_{21}/C_{22}) \sim 4 C_{21}/C_{22} $, and $E_{21}/E_{22} \propto
(\Omega_{22}/\Omega_{21})^2 (C_{21}/C_{22})^2 \sim 4
(C_{21}/C_{22})^2$, with $\Omega_{22}/\Omega_{21} \sim 2$ the ratio of
the frequencies of the two modes.

\begin{figure}
\raggedleft
\includegraphics[angle=90, clip=true,width=0.49\textwidth]{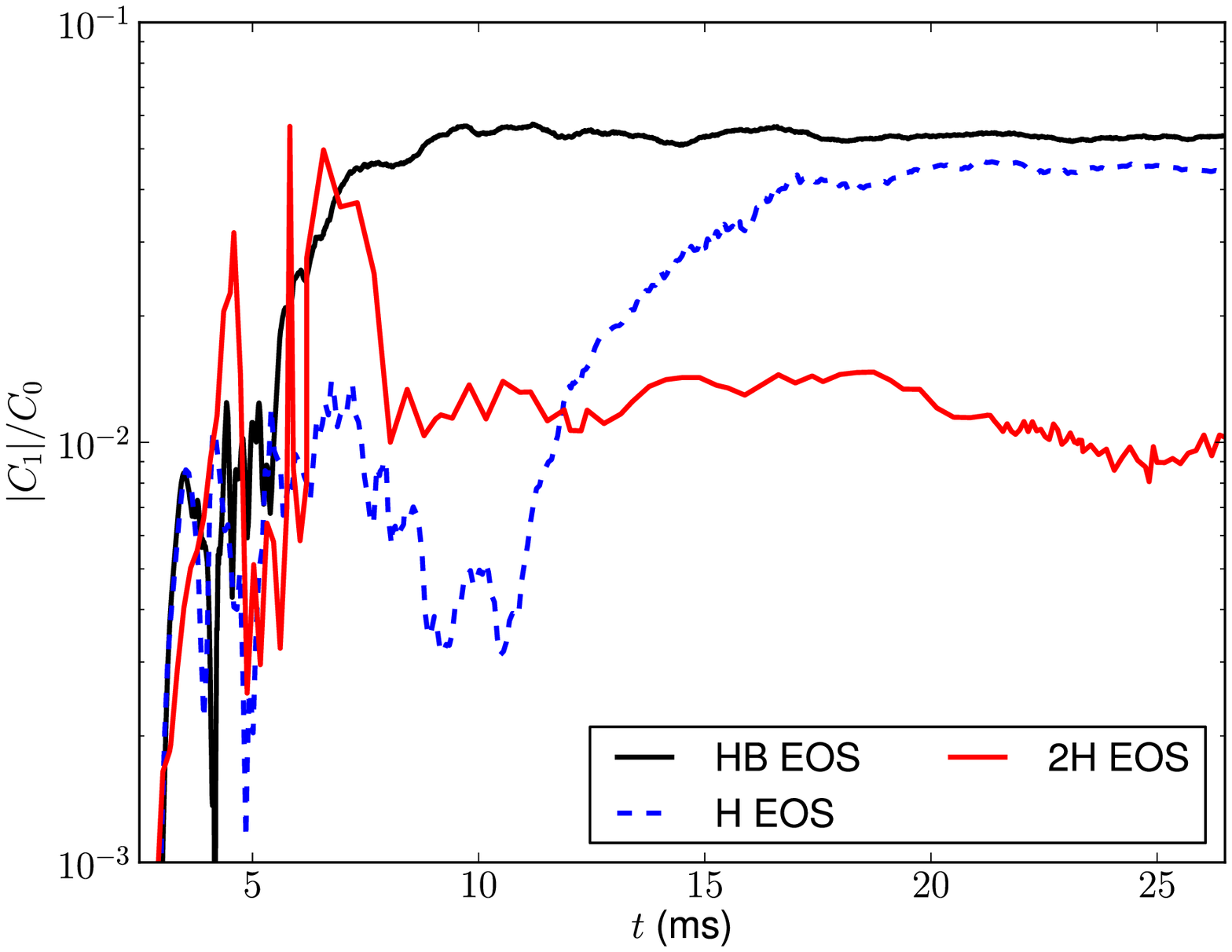}
\includegraphics[angle=90, clip=true,width=0.49\textwidth]{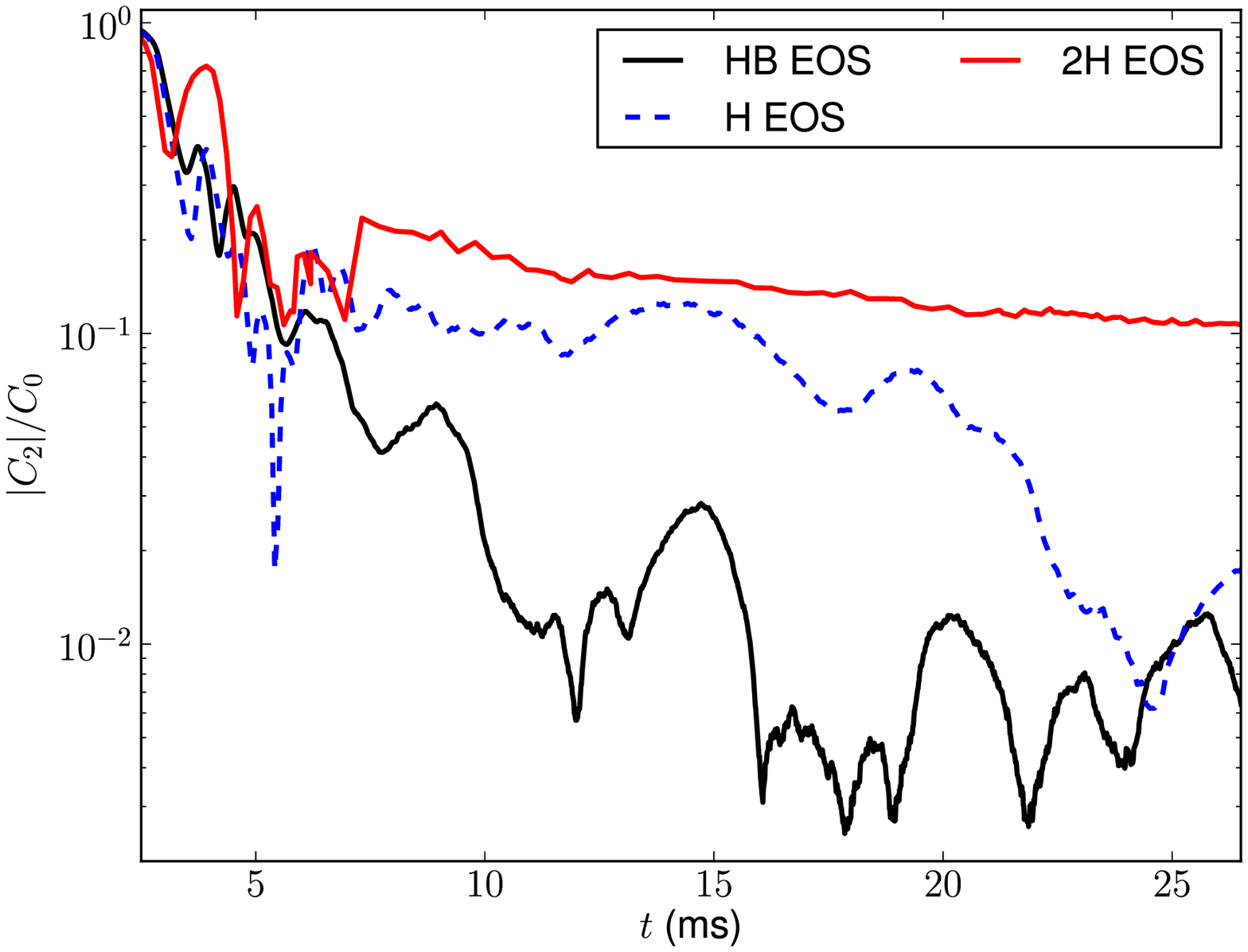}
\caption{
    A comparison of the amplitude of the $m=1$ (left) and $m=2$ (right) density 
    modes for equal-mass cases with $r_p/M=8$ and various EOSs. 
    }\label{m1m2_modes}
\end{figure}

In Fig.~\ref{m1m2_modes} we show a comparison of the $m=1$ and $m=2$
density modes for the different piece-wise polytropic EOSs that
produce long-lived HMNSs (i.e. all of them except the B EOS), again
focusing on equal-mass cases with the same impact parameter
$r_p=8M$. Here we can see that stiffer EOSs, which have more pressure
support, develop and maintain larger $m=2$ modes and correspondingly
smaller $m=1$ modes.  While $m=1$ modes eventually dominate over the
other nonzero density modes for the H and HB EOS, for the stiffest EOS
considered here, the 2H, after $\sim 35$ ms of evolution $C_2/C_1 \sim
2$. However, as we will discuss below, by considering larger impact
parameters or unequal masses, we do find cases with the 2H EOS where
the $m=1$ mode more strongly manifests.

In Table~\ref{modes_table} we list the dominant frequencies of the
$m=1$ and $m=2$ density modes that we obtain via a Fourier
transform of $C_1$ and $C_2$. As found before~\cite{PEPS2015,EPPS2016},
the frequencies satisfy $f_{m=2} \approx 2f_{m=1}$. Furthermore,
stiffer EOSs mimicking realistic ones seem to produce
lower frequency $m=1$ and $m=2$ modes. Although this is not as clean
in the azimuthal mode decomposition of the density (possibly because
of sparse sampling of the modes), it is more cleanly reflected in the
GW spectrum (see below).

%%%%%%%%%%%%%%%%%%%%%%%%%%%%%%%%%%%%%%%%%%%%%%%%%%%%%%%%%%%%%%%%%%%%%%%%%%%%
\begin{table}[t]
\caption{\label{modes_table} From left to right the columns correspond
  to the equation of state, the pericenter distance, the dimensionless
  spins of the two stars, the masses of the two stars in units of
  $M_\odot$, and the dominant frequencies of the $m=1$ ($f_{m=1}$) and
  $m=2$ ($f_{m=2}$) modes in units of kHz.  The last column indicates
  the approximate time after merger, in ms, at which the one-arm mode
  dominates the $m>1$ density modes (or a lower bound if this do not
  occur by the end of the simulated time).  Apart from the B equation
  of state, which collapses to a BH, all other cases form
  ``long-lived'' hypermassive neutron stars to which the results
  correspond.  We also include a lower resolution (2H,LR) run of one
  of our cases with the 2H EOS. }
\centering
\begin{tabular}{ccccccccc}
\hline\hline
EOS &
$ r_p$ &
$a_{\rm NS,1}$ &
$a_{\rm NS,2}$ &
$M_{\rm NS,1}$ &
$M_{\rm NS,2}$ &
$f_{m=1}$ & 
$f_{m=2}$ &
%$|C_1| > |C_{m>1}|$? 
$T_{\rm one-arm}$ 
\\
\hline
$k_1$  & 8M & 0.05 & 0.075 & 1.35 & 1.35 & 1.6 & 3.1 & 45 \\ % Yes \\
$k_2$  & 8M & 0.05 & 0.075 & 1.35 & 1.35 & 1.5 & 2.9 & 62 \\ % Yes \\
H      & 8M & 0.05 & 0.075 & 1.35 & 1.35 & 1.7 & 3.1 & 18 \\ % Yes \\
HB     & 8M & 0.05 & 0.075 & 1.35 & 1.35 & 1.7 & 3.3 & 6 \\ % Yes \\
2H     & 8M & 0.05 & 0.075 & 1.35 & 1.35 & 1.1 & 2.0 & $>28$ \\ % No  \\
HB     & 8M & 0.05 & 0.05  & 1.28 & 1.42 & 1.8 & 3.6 & 8\\ % Yes \\
2H     & 8M & 0.05 & 0.05  & 1.28 & 1.42 & 1.0 & 1.9 & 38\\ % Yes \\
2H,LR  & 8M & 0.05 & 0.05  & 1.28 & 1.42 & 1.0 & 2.0 & 12\\ % Yes \\
2H     &9.5M & 0.05 & 0.05 & 1.35 & 1.35 & 1.1 & 1.9 & 17 \\ % Yes \\
2H     & 8M & 0.05 & 0.05  & 1.70 & 1.70 & 1.1 & 2.3 & $>33$ \\ % No\V{?} \\
\hline\hline 
\end{tabular}
\end{table}
%%%%%%%%%%%%%%%%%%%%%%%%%%%%%%%%%%%%%%%%%%%%%%%%%%%%%%%%%%%%%%%%%%%%%%%%%%%%

\subsection{Double-core HMNSs}

\begin{figure}
\raggedleft 
\includegraphics[trim =2.0cm 0.0cm 2.0cm  0.cm,clip=true,width=0.49\textwidth]{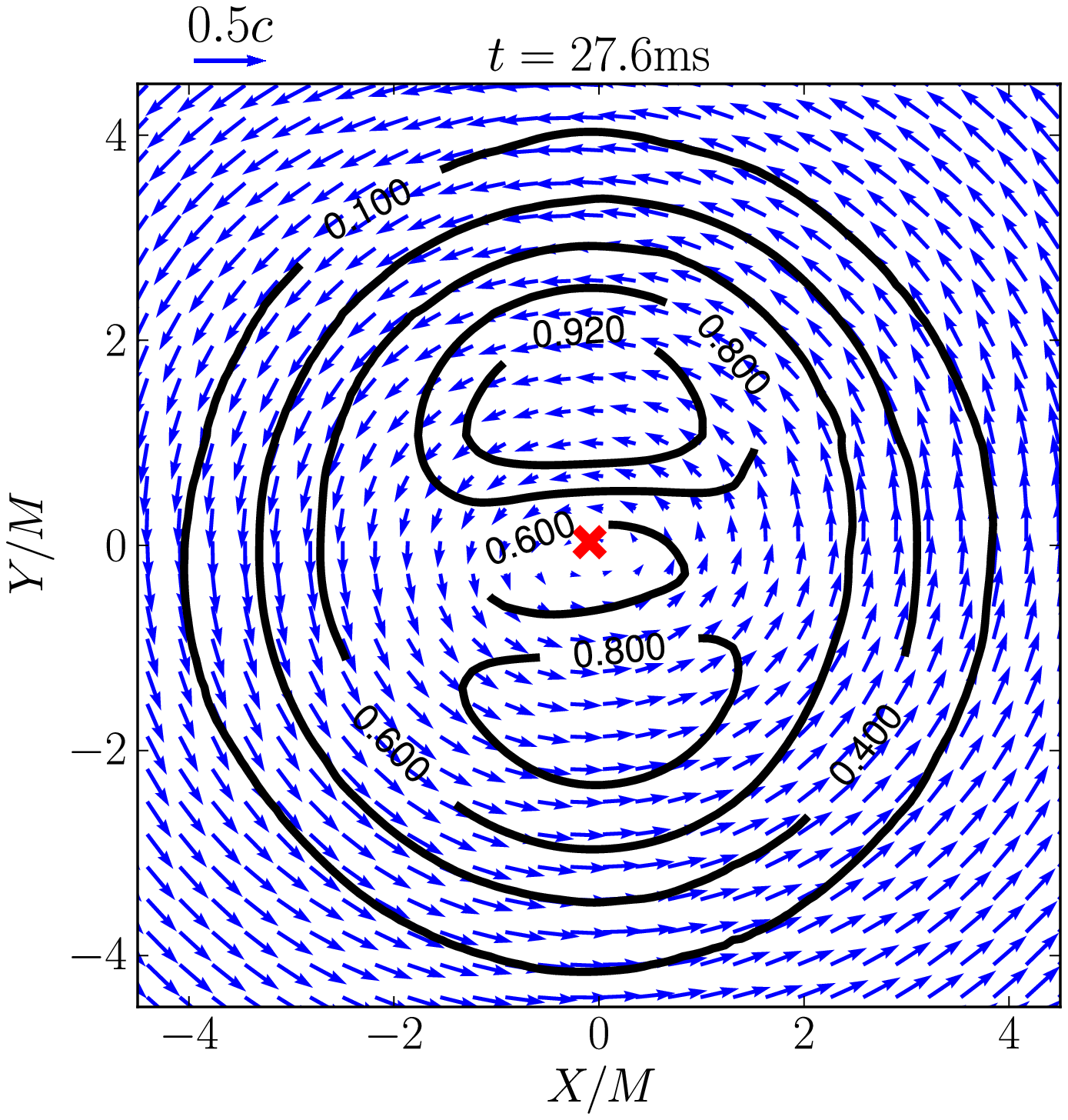}
\includegraphics[trim =2.0cm 0.0cm 2.0cm  0.cm,clip=true,width=0.49\textwidth]{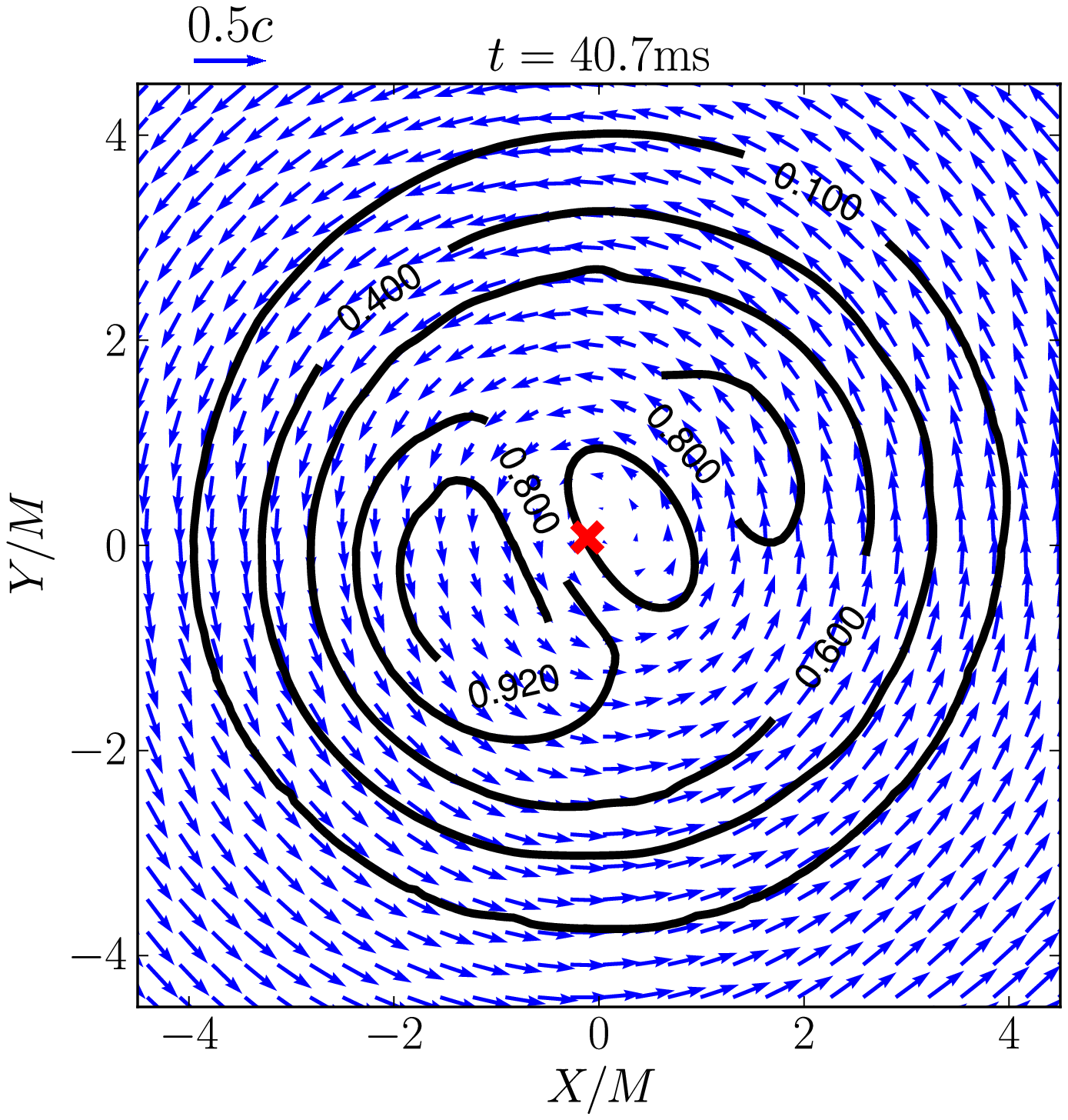}
\includegraphics[trim =2.0cm 0.0cm 2.0cm  0.cm,width=0.49\textwidth]{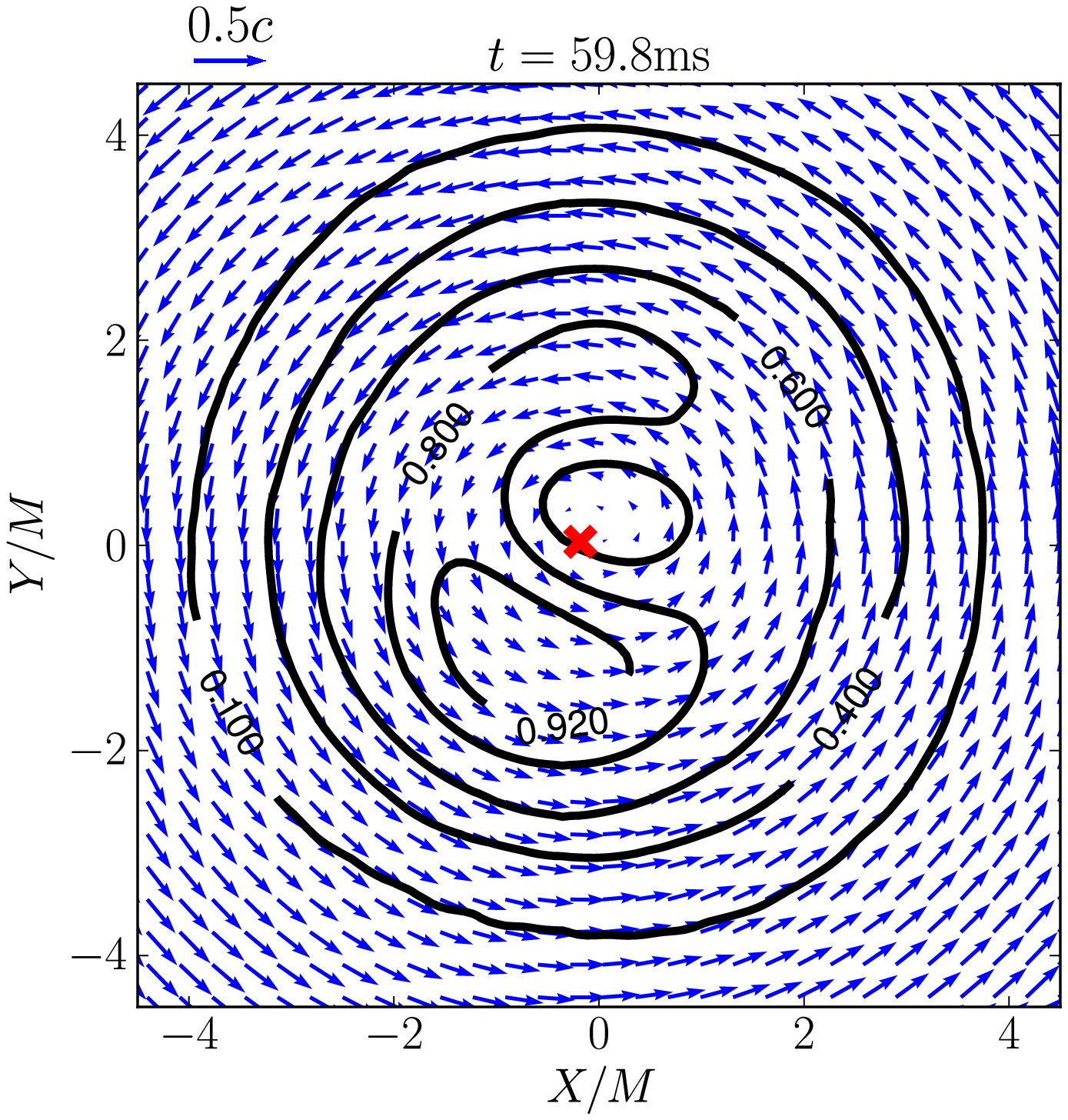}
\includegraphics[trim =2.0cm 0.0cm 2.0cm  0.cm,width=0.49\textwidth]{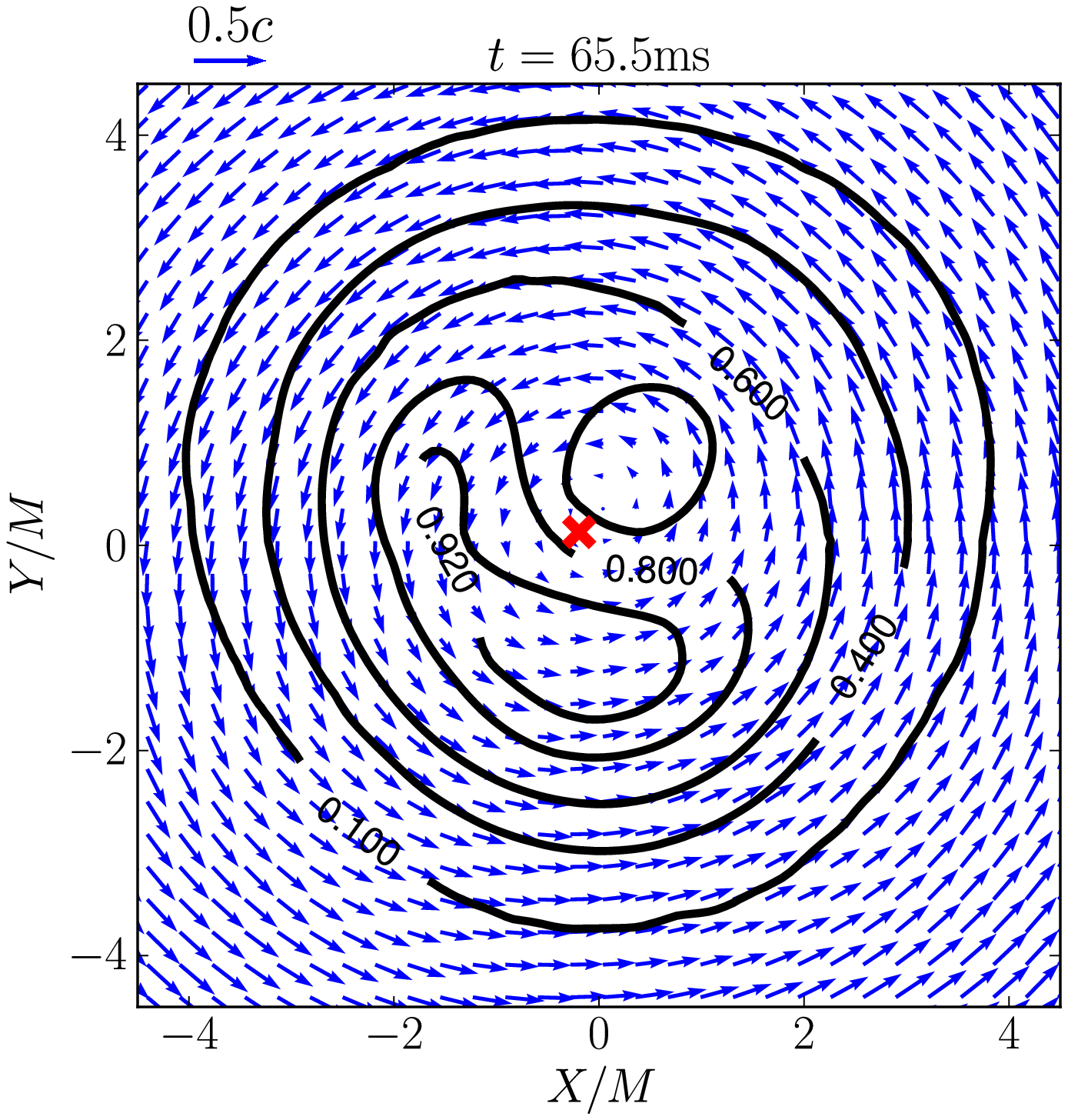}
\includegraphics[trim =2.0cm 0.0cm 2.0cm  0.cm,width=0.49\textwidth]{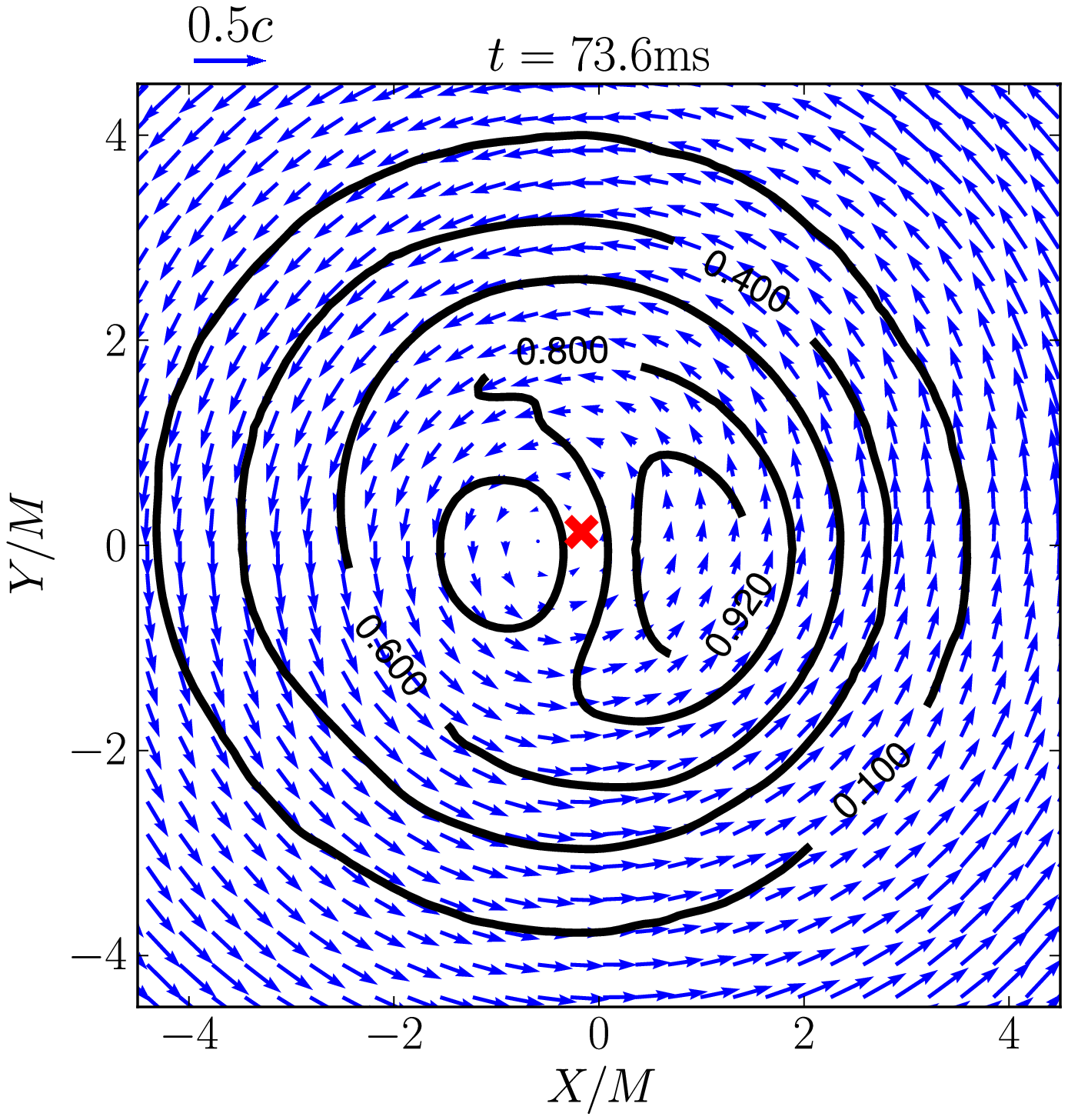}
\includegraphics[trim =2.0cm 0.0cm 2.0cm  0.cm,width=0.49\textwidth]{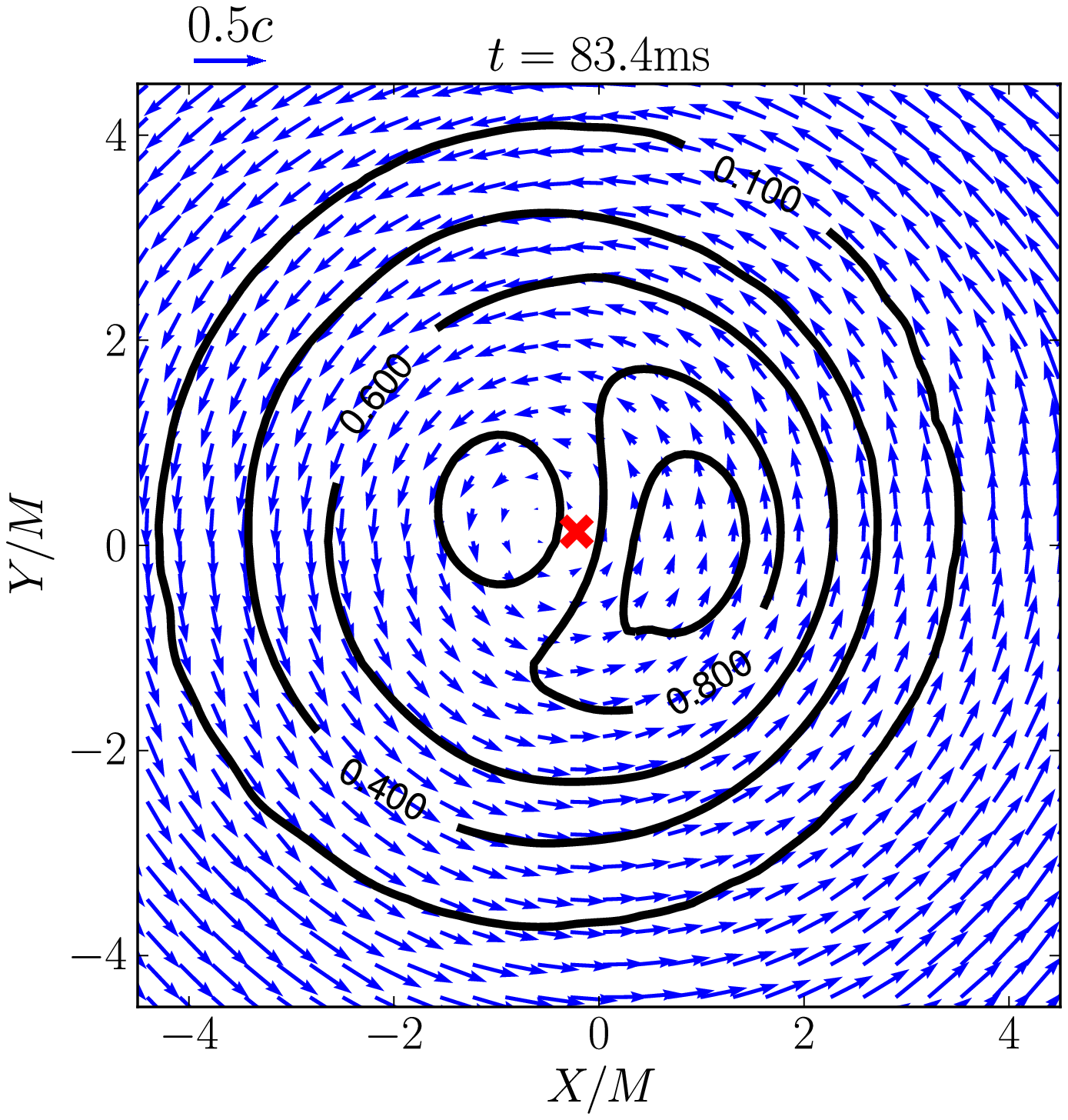}
\caption{Contours of rest-mass density normalized to its maximum value
  (black solid lines) and coordinate velocity (arrows) on the
  equatorial plane at select times following merger. The (red) ``x''
  indicates the HMNS center of mass. The data in all plots are taken
  from the $r_p/M=8$, $a_{\rm NS,1}=0.05, a_{\rm NS,2}=0.075$, $k_2$ case.
}\label{vec2Dgam3}
\end{figure}

For the cases with the $k_1$ and $k_2$ EOSs, we find that a
double-core HMNS forms post-merger, i.e. there are two distinct maxima
in the density of the star. This is illustrated in
Fig.~\ref{vec2Dgam3} where we show snapshots of density contours and
the velocity flow from the evolution of the $r_p=8M$, $a_{\rm
  NS,1}=0.05, a_{\rm NS,2}=0.075$, $k_1$ case.  In particular,
following merger the two dense cores rotate around the common center
of mass, which is surrounded by an underdense region (upper row in
Fig.~\ref{vec2Dgam3}). One of the two cores is slightly larger than
the other due to the initial $m=1$ asymmetry. The subsequent evolution
is such that matter from the smaller core gradually flows towards that
of the larger core until eventually the two cores merge into one. At
this time the $m=1$ instability is clearly visible in the $C_m$ modes
in that $|C_1| > |C_m|, \ m=2$, 3, and 4.

In previous works, we reported that the background about which the
one-arm instability develops was a toroidal-like HMNS (maximum density
occurring in a ring around the center of mass on the equatorial
plane), that formed following the interaction of post-merger fluid
vortices arising at the shear layers near the surface of stars and the
tidal tails during merger. Here, for many of the piece-wise polytropic
EOS cases we do not find the same vortex dynamics as in our previous
work, and the HMNS instead forms an ellipsoidal configuration, but is
still subject to the $m=1$ instability. This is also the case for
quasi-circular binary NS mergers reported
in~\cite{2016arXiv160305726R}.  For the $k_1$ and $k_2$ polytropic EOS
we find a third background: a double-core HMNS.  This suggests that
such details of the structure of the remnant do not play a significant
role in determining whether conditions are ripe for eventual growth of
the $m=1$ mode.

In the double core case, an asymmetry in the mass of the two cores
causes a large $m=1$ mode to be present well before the cores merge
and the one arm mode manifests. This is then another way
strong $l=2$, $m=1$ gravitational-wave modes can be generated. 
Thus, detection of a near quasistationary $l=2$, $m=1$ GW signal may not
necessarily imply the one-arm spiral instability. However, the results
here suggest this degeneracy could be broken if one folds in information from the
$l=2$, $m=2$ GW modes, which appear to be much stronger than the $l=2$,
$m=1$ GW modes during the double core phase.

The $\Gamma=3$ polytropic stars we consider as our initial data here
are evolved with a $\Gamma$-law EOS. This implies that $\Gamma_{\rm
  th}=3$. While this choice of $\Gamma_{\rm th}$ is large compared to
what is typically adopted in the literature, it likely does not affect
the physics of high-density matter that gives rise to the
double cores, whose densities are approximately the same as the
initial maximum density in the stars. The reason is that strong
shock heating here (and hence high thermal pressure) occurs only in low
density matter. For example using Eq. B6 of Appendix B
in~\cite{Etienne2009PhRvD} and setting $\Gamma=3$ with $k\rho^2 \sim
0.4^2$ (which holds for our $\Gamma=3$ models), one finds that in
the strong shock limit the ratio of total pressure to cold pressure
for our models is $K \sim 0.06(v/0.4)^2 (\rho_{0,\rm max}/\rho_0)^2$,
where $v$ is the collision velocity. Thus, only for rest-mass
densities $\rho_0 < 0.1\rho_{0,\rm max}$ can there be a significant
shock heating effect. But, since a $\Gamma=3$ EOS is stiff the bulk of
the total neutron-star mass lies in the density range
$\rho_0 > 0.1\rho_{0,\rm max}$.

While our limited set of simulations cannot rule out the
possibility that the double-core remnant arises due to the high value
of $\Gamma_{\rm th}$, we point out that double-core HMNS remnants are
known to form even for lower values $\Gamma_{\rm
  th}=1.357$~\cite{2011PhRvD..83l4008H}.  Thus, the results from our
simulations employing a $\Gamma$-law with $\Gamma=3$ should simply be
viewed as a qualitative representation of a class of EOSs for which
binary neutron star mergers lead to the formation of double core
HMNSs.%~\cite{2011PhRvD..83l4008H}.

\subsection{Effects of varying mass ratio and other parameters}

\begin{figure}
\raggedleft
\includegraphics[trim =0.25cm 0.2cm 2.5cm  0.2cm,clip=true,width=0.49\textwidth]{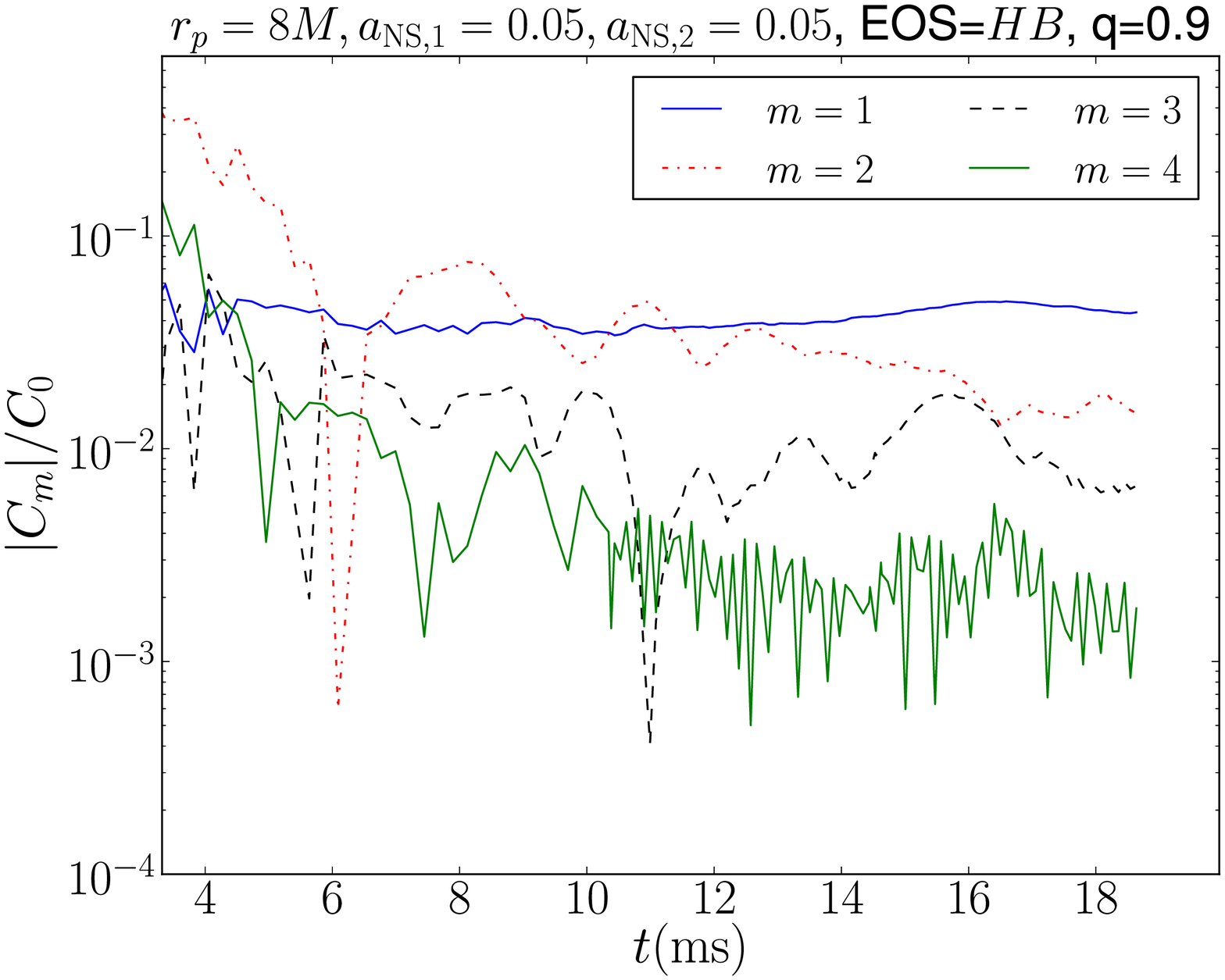}
\includegraphics[trim =0.25cm 0.2cm 2.5cm  0.2cm,clip=true,width=0.49\textwidth]{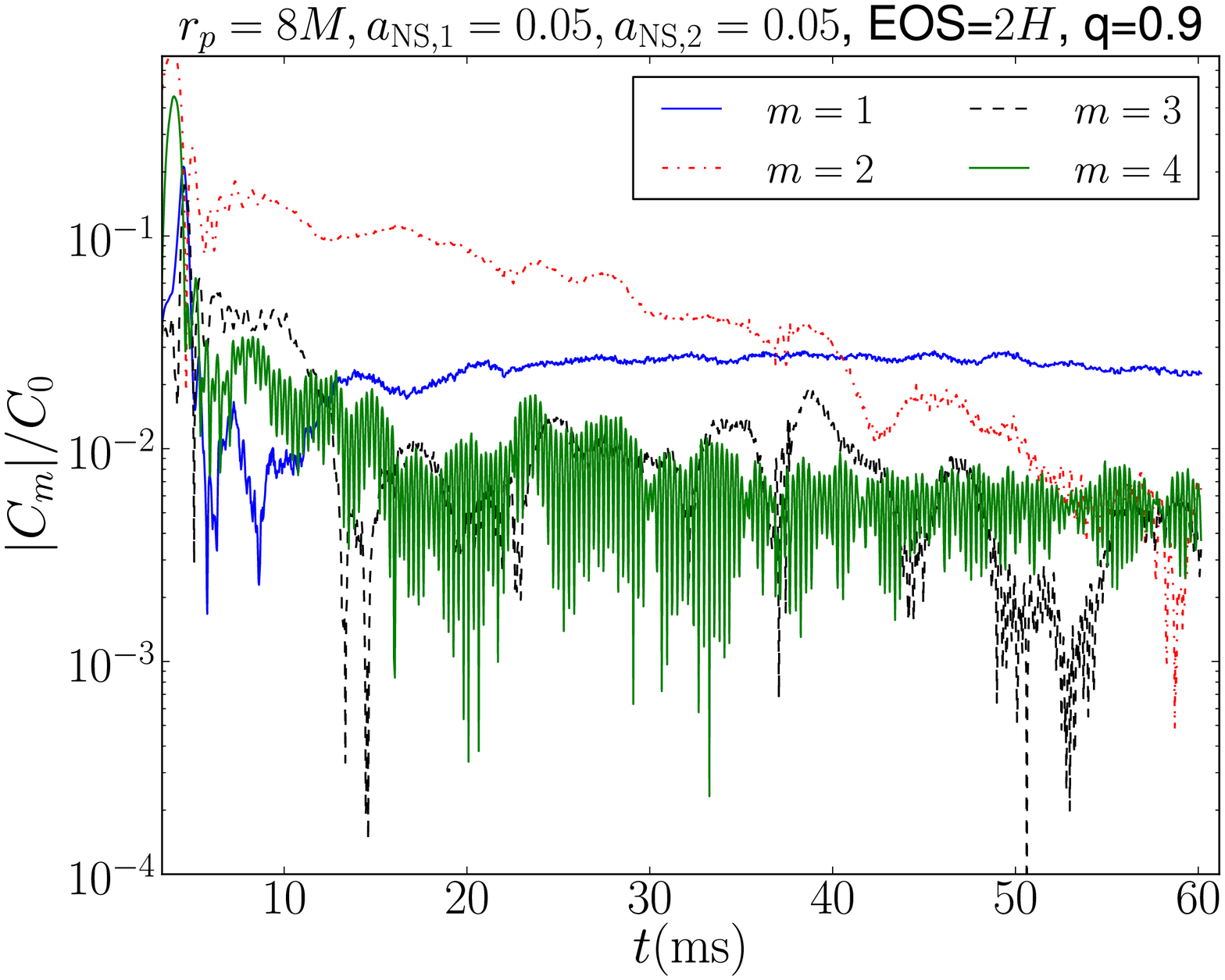}
\caption{Amplitude of $C_m$ normalized to $C_0$ for $q=0.9$
  cases (note the different time ranges for each plot).
}\label{modesq09}
\end{figure}

\begin{figure}
\raggedleft
\includegraphics[trim =0.25cm 0.2cm 0.5cm  0.2cm,clip=true,width=0.49\textwidth]{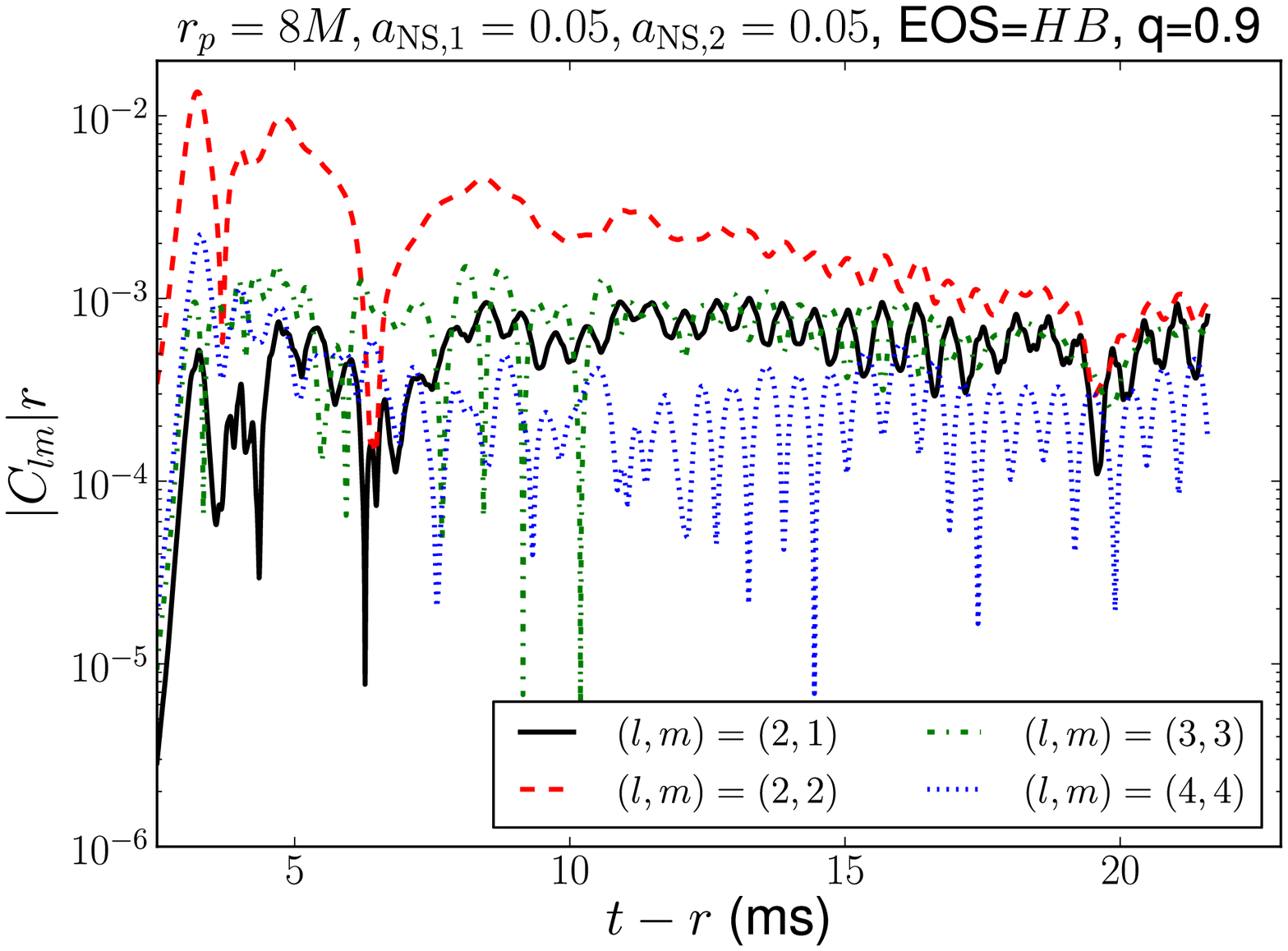}
\includegraphics[trim =0.25cm 0.2cm 0.5cm  0.2cm,clip=true,width=0.49\textwidth]{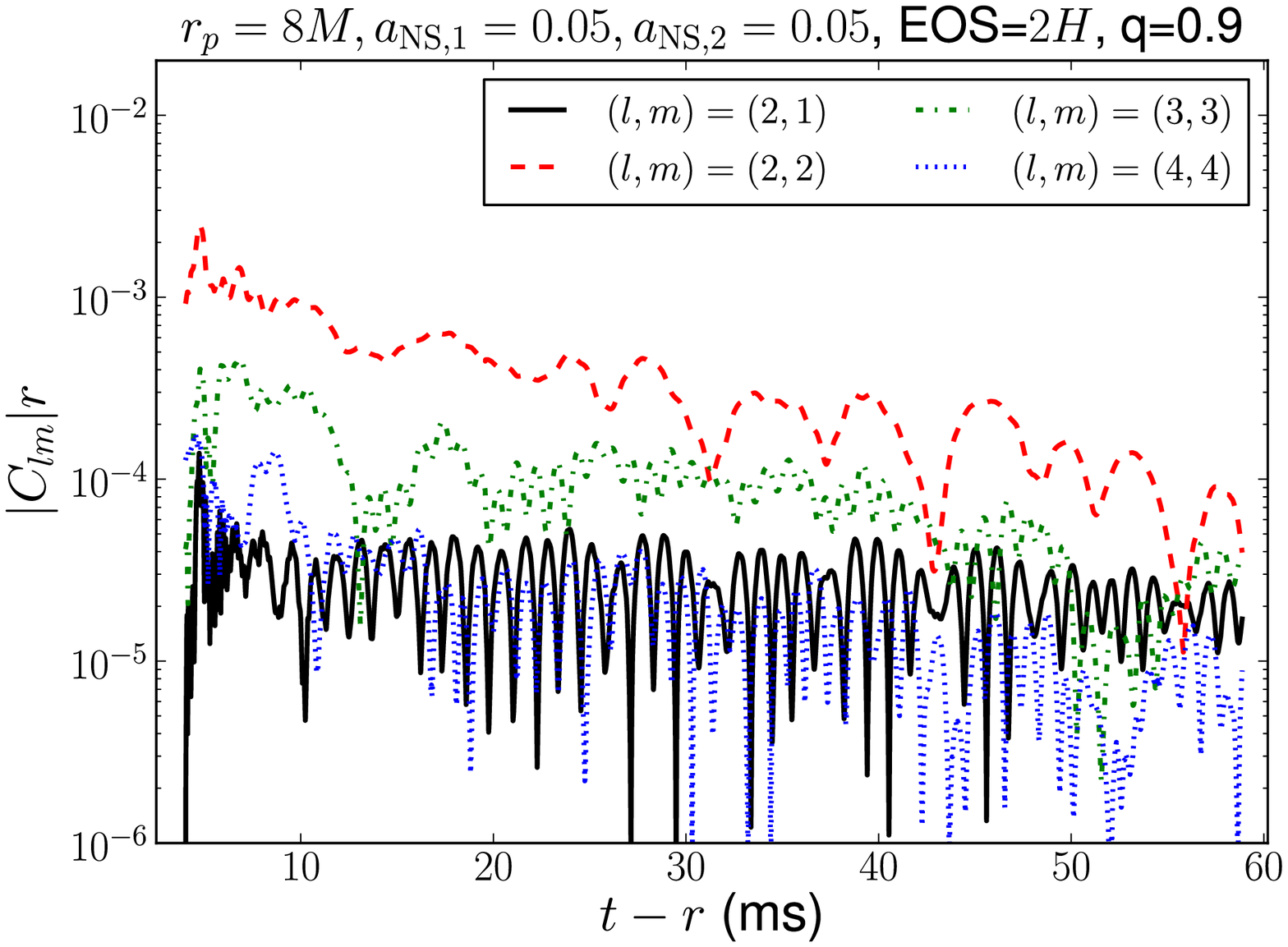}
\caption{Amplitude of several spherical harmonic components of the GW
  signal, normalized by $M$, following merger for the same cases shown in corresponding panels
  in Fig.~\ref{modesq09}.}\label{gwclmEOSstudy_q09}
\end{figure}

In addition to the equal mass cases considered above, we also consider
two cases with mass ratio $q=0.9$.  We use two equations of state, the
HB and 2H, covering a range of stiffness from moderate to very stiff
EOS, while keeping the total gravitational mass fixed to $2.7M_\odot$,
the periapse distance fixed to $r_p = 8M$, and the individual NS
dimensionless spins fixed to $0.05$.  By virtue of the unequal
mass-ratio, there is a strong $m=1$ component to the initial data. We
find that the $m=1$ instability occurs in both cases.  In
Fig.~\ref{modesq09} we plot the amplitude of $|C_m|$ as a function of
time demonstrating the dominance of the $m=1$ azimuthal modes. The
plots demonstrate that soon after merger the $m=1$ mode reaches
saturation, while the $m=2$ mode decays and eventually its amplitude drops below that of the
$m=1$ mode, at which point the $m=1$ structure is clearly seen in 2D equatorial
density plots.  The corresponding decomposition of the GWs is shown in
Fig.~\ref{gwclmEOSstudy_q09}.

\subsubsection{Varying other parameters with the 2H EOS:}
\begin{figure}
\raggedleft
\includegraphics[angle=90, clip=true,width=0.49\textwidth]{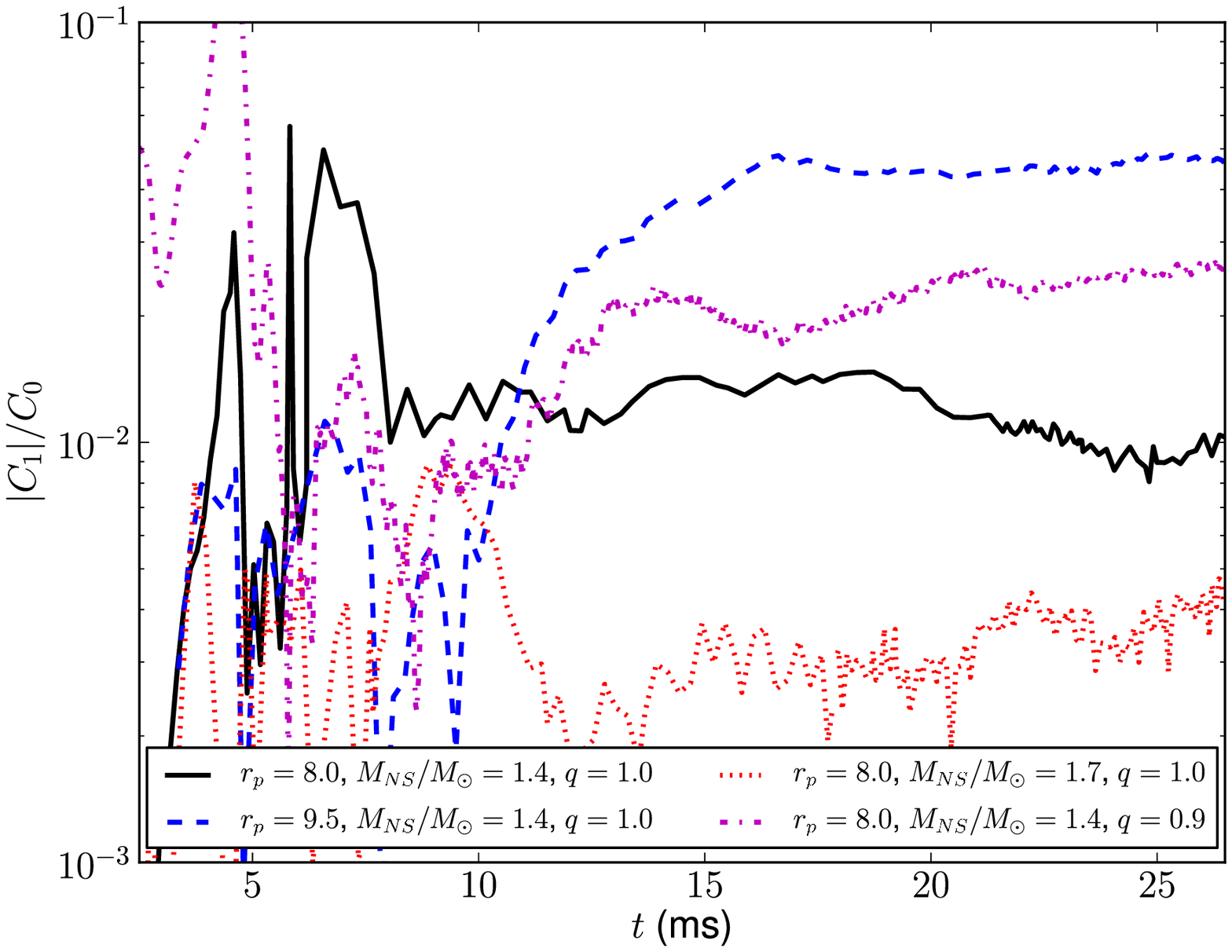}
\includegraphics[angle=90, clip=true,width=0.49\textwidth]{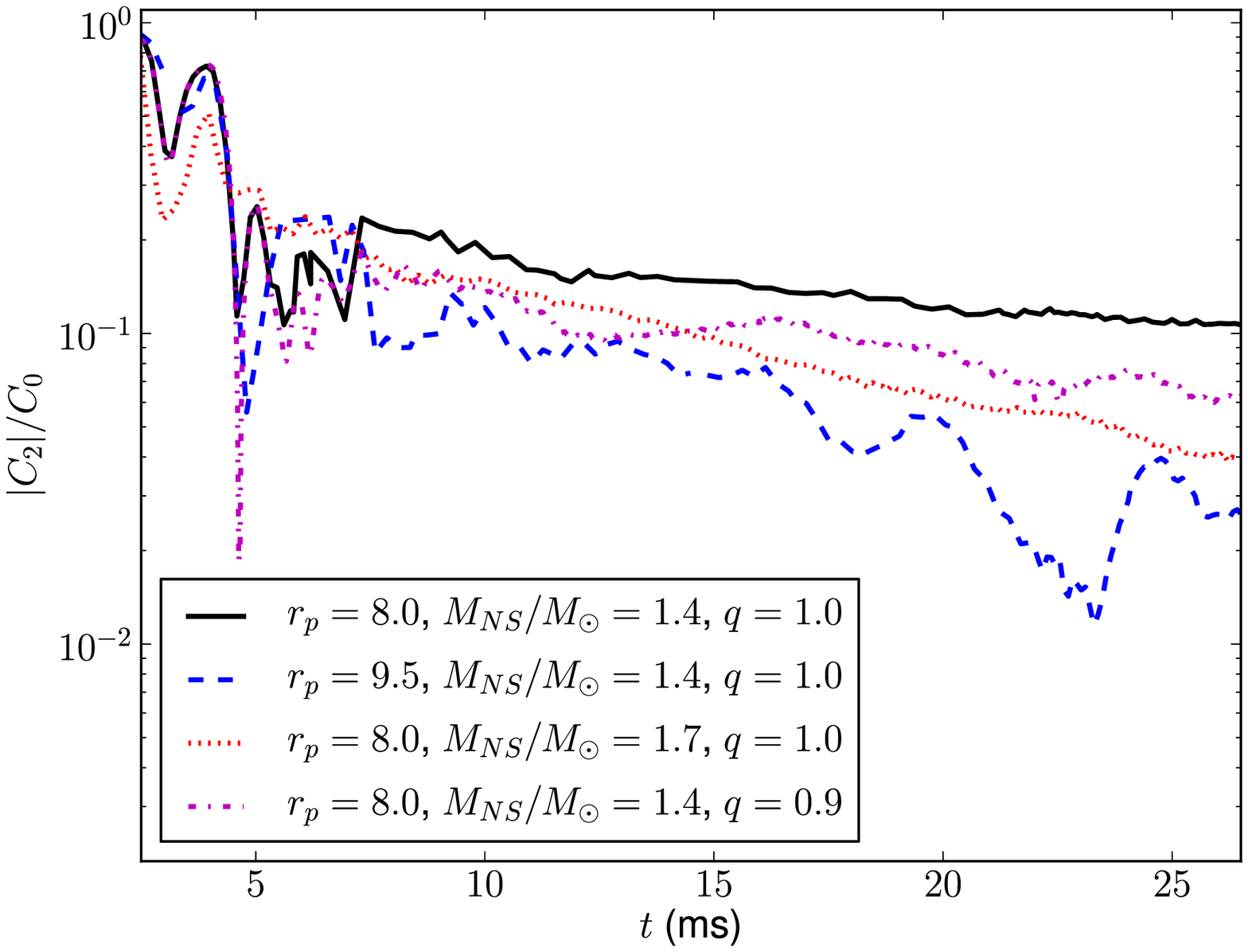}
\caption{
    A comparison of the amplitude of the $m=1$ (left) and $m=2$ (right) density 
    modes for various cases with the 2H EOS.}\label{m1m2_modes_2H}
\end{figure}

Above we showed that while the equal-mass 2H EOS case exhibited a
strong $m=2$ mode and suppressed $m=1$ mode in the post-merger HMNS,
the $m=1$ mode was noticeably enhanced in the $q=0.9$ case.  To probe
the effects of the other binary parameters, we also consider
equal-mass cases with this EOS where we increase the impact parameter
to $r_p=9.5M$, and hence the angular momentum at merger; we further
examined a case where we where we
increase the total mass of each NS from 1.35 to 1.7 $M_{\odot}$,
giving more compact stars.  The $m=1$ and $m=2$ density modes from all
four cases with the 2H EOS are compared in Fig.~\ref{m1m2_modes_2H}.
The increased impact parameter does noticeably enhance the strength of
the one-arm mode.  One the other hand, the case with higher mass stars
shows a significantly smaller $m=1$ mode.  We find that when
evolved to $\approx 36$ ms, the $m=2$ mode decays noticeably, though
it still dominates over the $m=1$ mode by roughly a factor of two.

Finally, based on the results listed in
Table~\ref{modes_table}, we conclude that for a given total mass
there is little variation ($\sim 10\%$) of the mode frequencies when
varying the mass ratio, the NS spins and the pericenter distance in
the 2H EOS. The approximate independence of the mode frequency with
the mass ratio has also been reported
in~\cite{2016arXiv160502369L}. Given that the density-mode frequencies
are reflected in the gravitational wave spectrum (see below), the small
variation of the GW frequencies with these different binary parameters spells good
news for the ability to constrain the nuclear EOS with future GW
observations, as the HMNS oscillations seem to
be largely determined by the total mass and the equation of state (at least
within the context of our high-eccentricity mergers). However,
detailed investigations are required to draw robust conclusions and
derive high-accuracy results that can be useful for future GW
observations. Such investigations will be the subject of future work
of ours.

\subsubsection{Effects of resolution:}
\begin{figure}
\raggedleft
\includegraphics[trim =0.25cm 0.2cm 2.5cm  0.2cm,clip=true,width=0.46\textwidth]{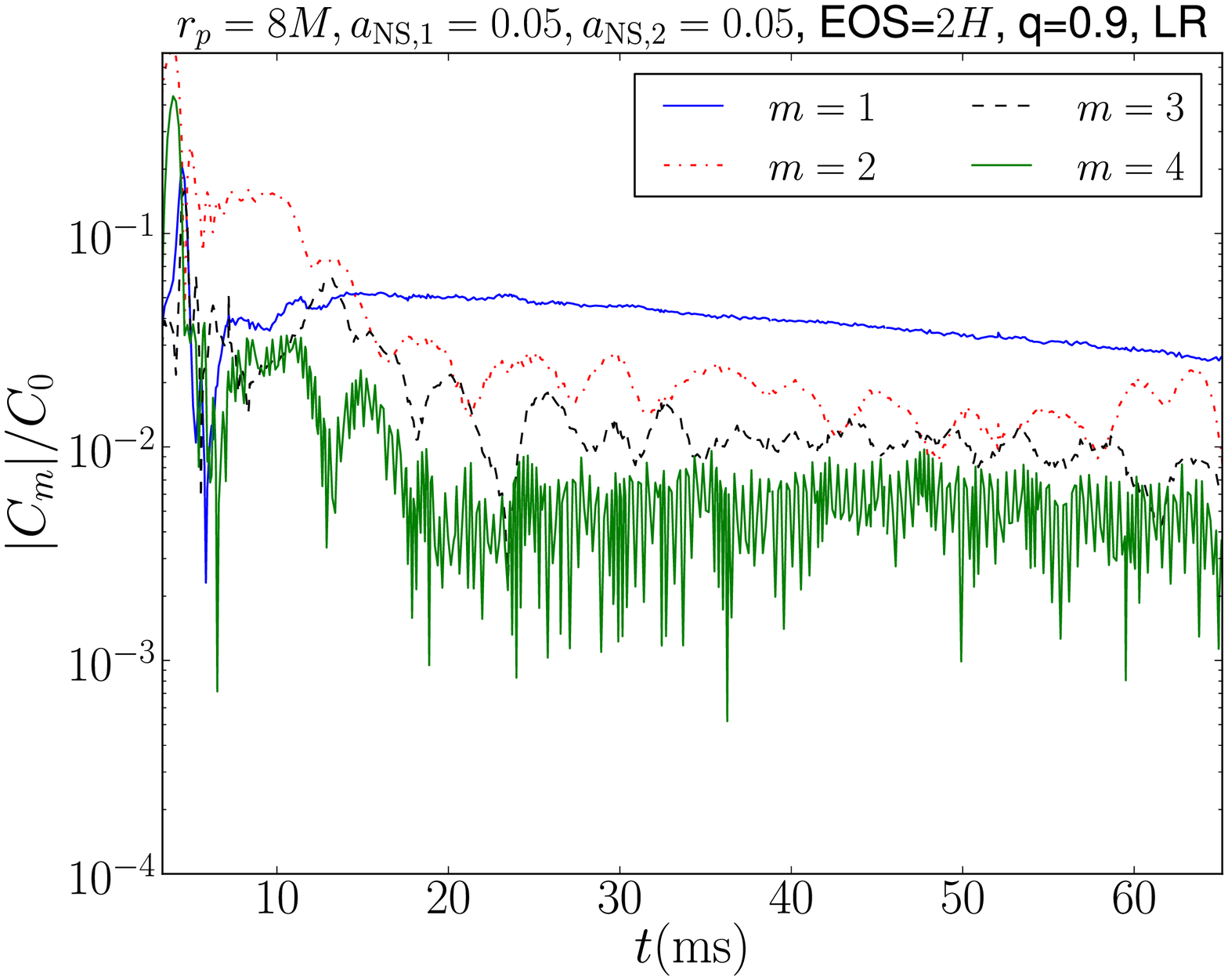}
\includegraphics[trim =0.25cm 0.75cm 0.5cm  0.65cm,clip=true,width=0.515\textwidth]{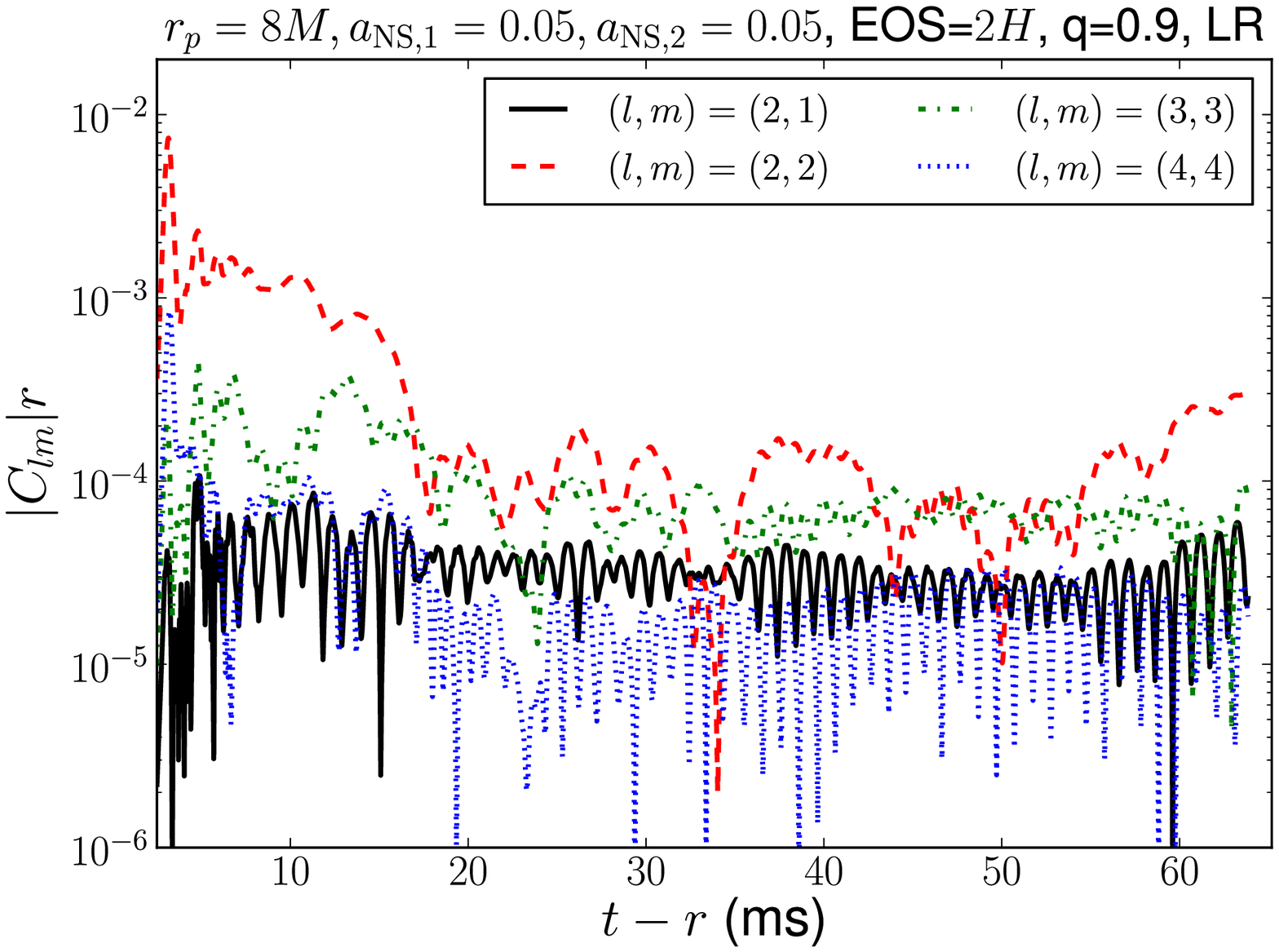}
\caption{ Results from a lower resolution run of the $q=0.9$, 2H EOS case.
  Left: Amplitude of $C_m$ normalized to $C_0$ (compare right panel
  of Fig. ~\ref{modesq09}). Right: Amplitude
  of several spherical harmonic components of the GW signal following
  merger (compare right panel of Fig.~\ref{gwclmEOSstudy_q09}) .}\label{modesGWsq09LR}
\end{figure}

Resolution studies demonstrating the convergence properties of the
solution for HB EOS cases were presented in ~\cite{EPPS2016}. Here we
supplement this with an additional 2H EOS, $q=0.9$ case run at a
resolution $0.64$ times lower than the resolution for the other
results above.  The density mode analysis and the GWs from this case
are shown in Fig.~\ref{modesGWsq09LR}.  The most notable feature in
the lower resolution run is that the $m=2$ density modes decay much
faster than in the higher resolution run, as does the $l=2,\ m=2$
mode of the GWs. As a result the $m=1$ instability manifests earlier
in the lower resolution run. These results indicate that for accurate
computation of all quantitative hydrodynamic effects higher resolution
would need to be adopted; however, the appearance and qualitative
nature of the $m=1$ instability, such as the approximately constant
amplitude of the $l=2,\ m=1$ GW mode, appears relatively insensitive
to resolution. Moreover, the frequencies of the density modes (see
Table~\ref{modes_table}) agree to within 5\% between the two different
resolutions we adopt in the 2H $q=0.9$ case.

\subsection{Gravitational wave spectrum}

\begin{figure}
\raggedleft
\includegraphics[angle=90, clip=true,width=0.49\textwidth]{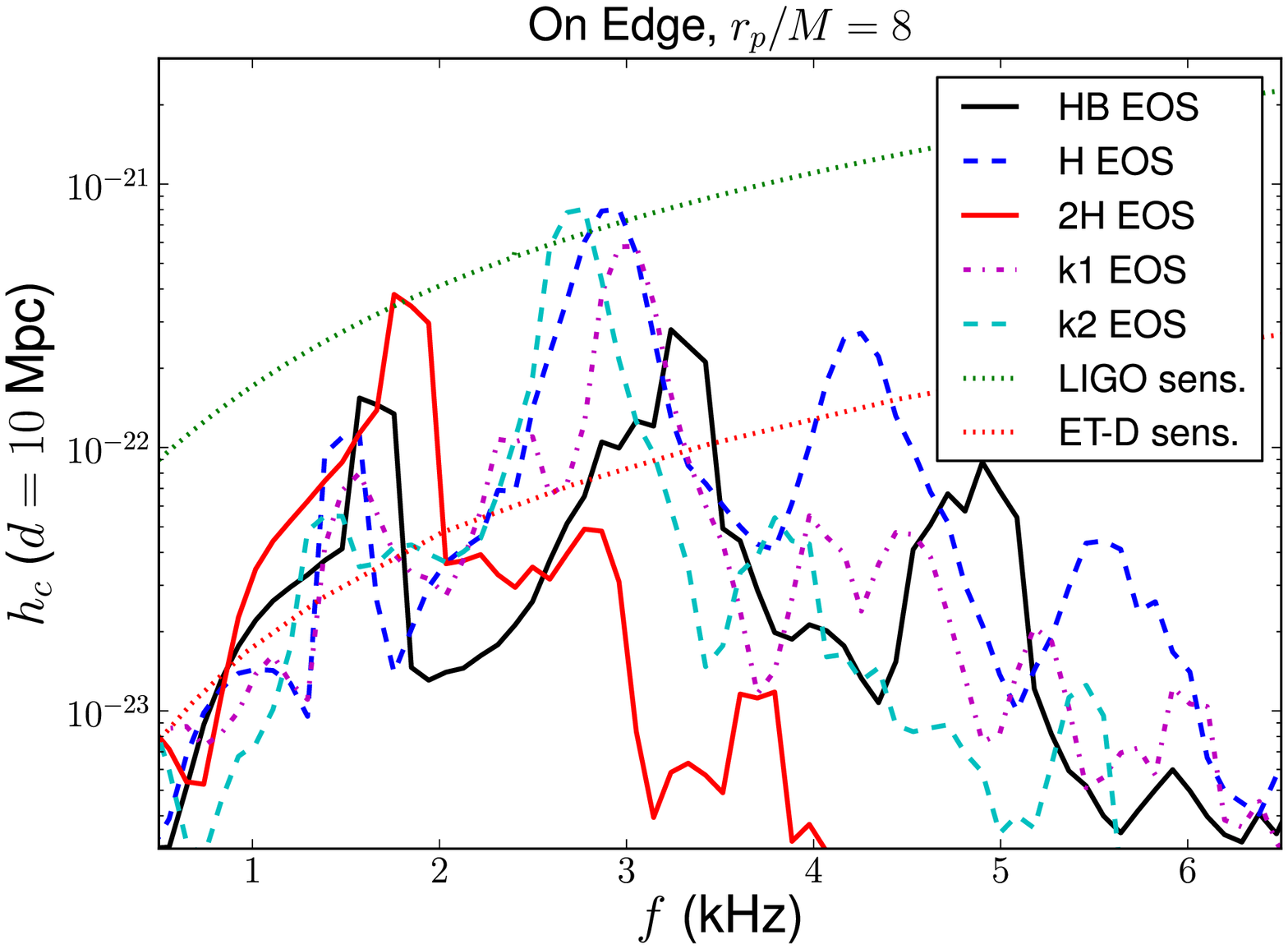}
\includegraphics[angle=90, clip=true,width=0.49\textwidth]{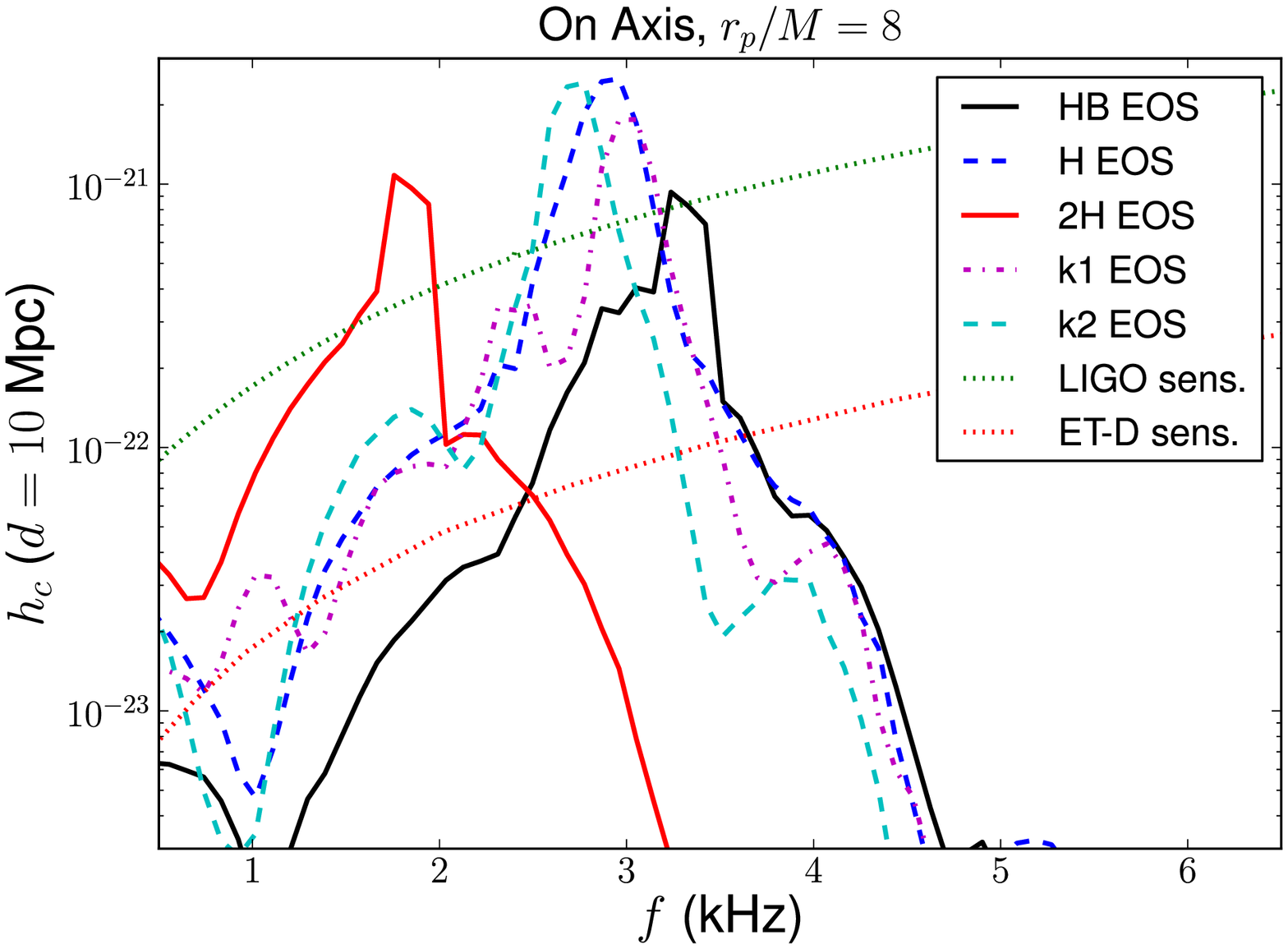}
\includegraphics[angle=90, clip=true,width=0.49\textwidth]{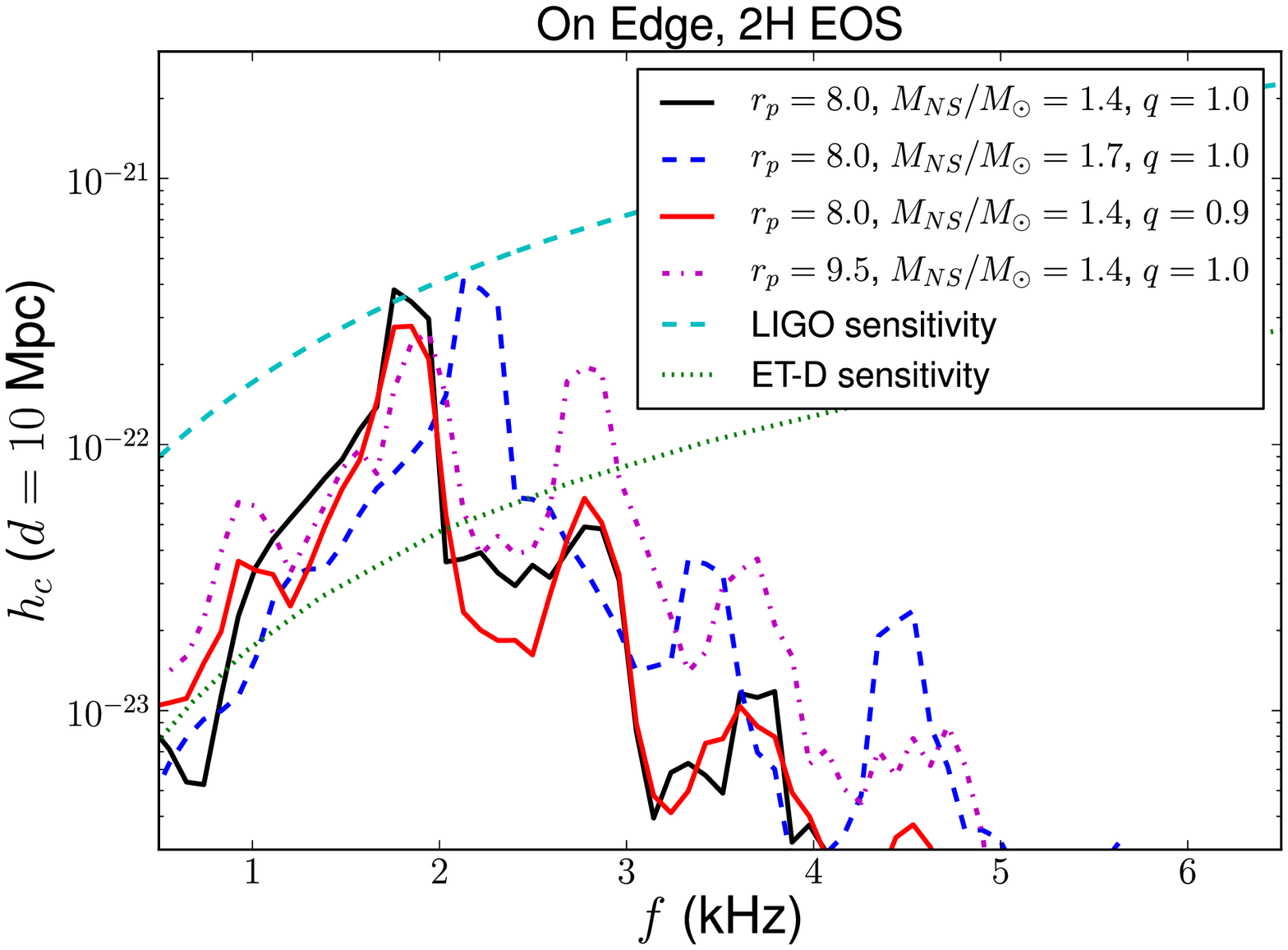}
\includegraphics[angle=90, clip=true,width=0.49\textwidth]{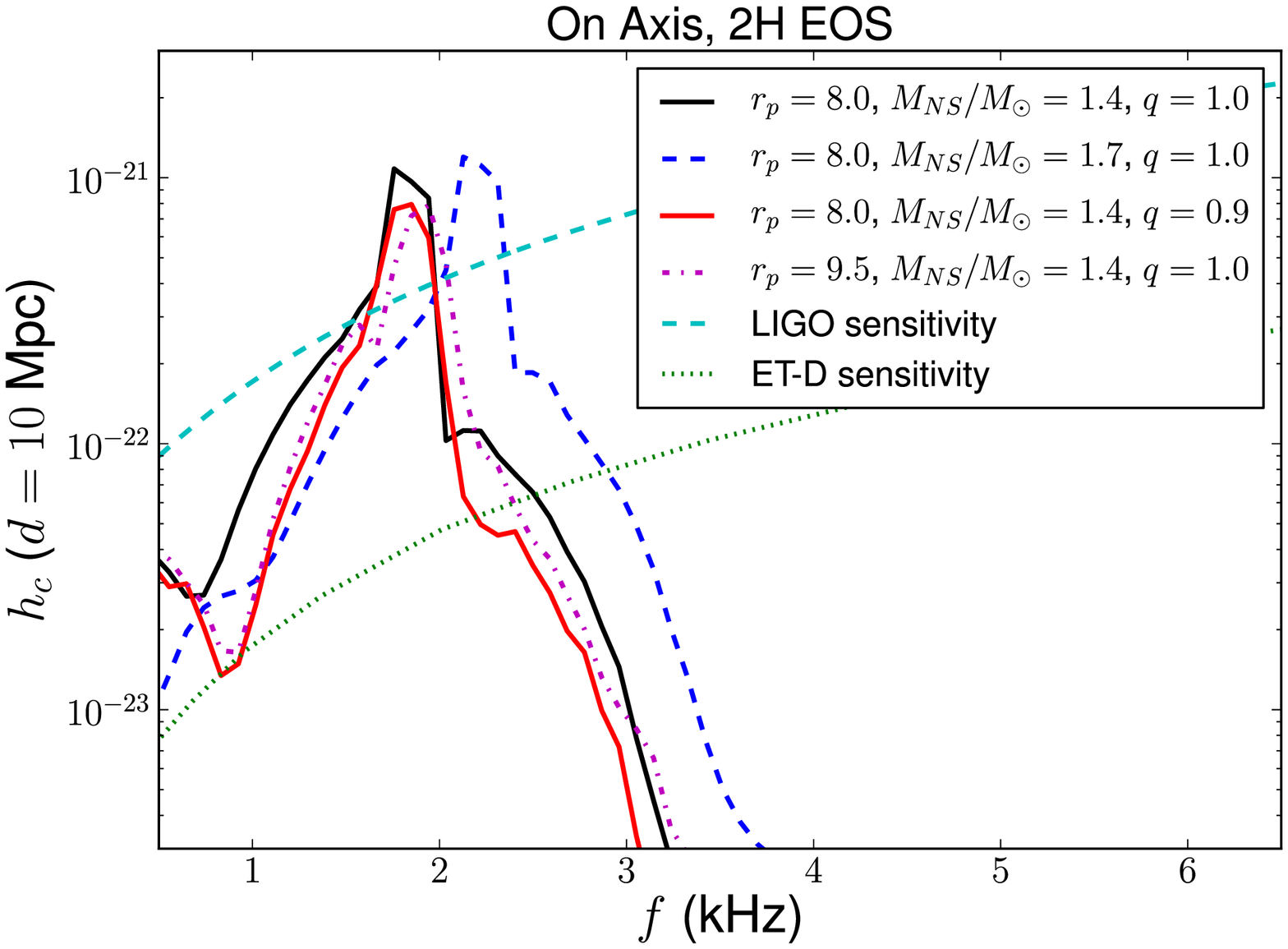}
\caption{
    The characteristic gravitational wave strain as a function of frequency
    for an observer on edge (left panels) and on axis (right panels)
    for $\approx 10$ ms of the post-merger phase.
    The top panels show equal-mass cases with $r_p/M=8$ and various EOSs while 
    the bottom panels show various cases with the 2H EOS. We also show the
    sensitivity curves for the aLIGO and Einstein Telescope (ET) detectors.
    }\label{gwSpec}
\end{figure}

The properties of the oscillating HMNS are encoded in the
gravitational wave spectrum post-merger.  In Fig.~\ref{gwSpec} we plot
the characteristic GW strain $h_c=|\tilde{h}|f$, defined in terms of
the Fourier transform $\tilde{h}$ of the strain and frequency $f$. We
show an integration of 10 ms post-merger for a consistent comparison
amongst the various cases. In Fig.~\ref{gwSpec} we show the strain
both as it would be measured by an observer on axis, where the
$(l,m)=(2,2)$ will be maximal and the $(2,\pm1)$ modes will vanish,
and by an observer on edge. In particular, in the on-edge plots a
series of peaks can be seen at integer multiples of the frequency of
the oscillation of the NS.  The strongest peak corresponds to the
dominant $(l,m)=(2,2)$ mode (at roughly 2 kHz for the 2H EOS, and
$\sim 3$ kHz for the softer EOSs), with the subdominant $(2,1)$ mode
at roughly half the $(2,2)$ mode's frequency, hence in a more
sensitive regime for GW detectors.  As expected, the cases with softer
EOSs, and thus more compact stars, have higher frequency peaks.  Likewise
the one case where the two NSs have a larger mass of $1.7$ $M_{\rm
  odot}$ has a higher frequency compared to the equivalent case where
the NSs have a mass of 1.35 $M_{\rm odot}$.  Again, the fact that the
stiffer EOSs have weaker GWs coming from the $m=1$ oscillations is
somewhat offset by the fact that the sensitivity is higher at lower
frequencies.  For fixed total mass, changing the mass-ratio or the
impact parameter has a relatively small effect on the frequency of the
GWs emitted in this time window after merger.

\begin{figure}
\raggedleft
\includegraphics[clip=true,width=0.49\textwidth]{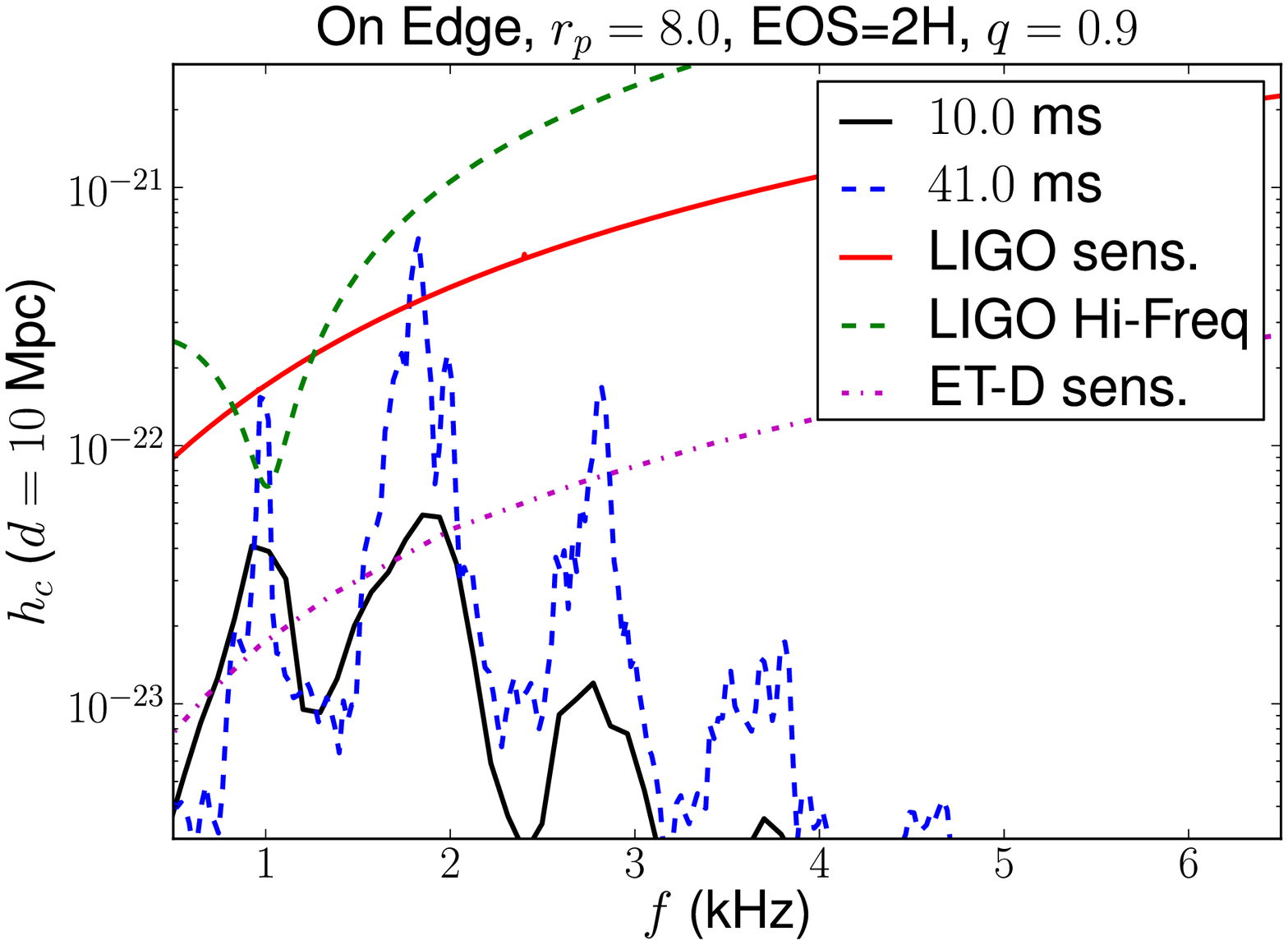}
\includegraphics[clip=true,width=0.49\textwidth]{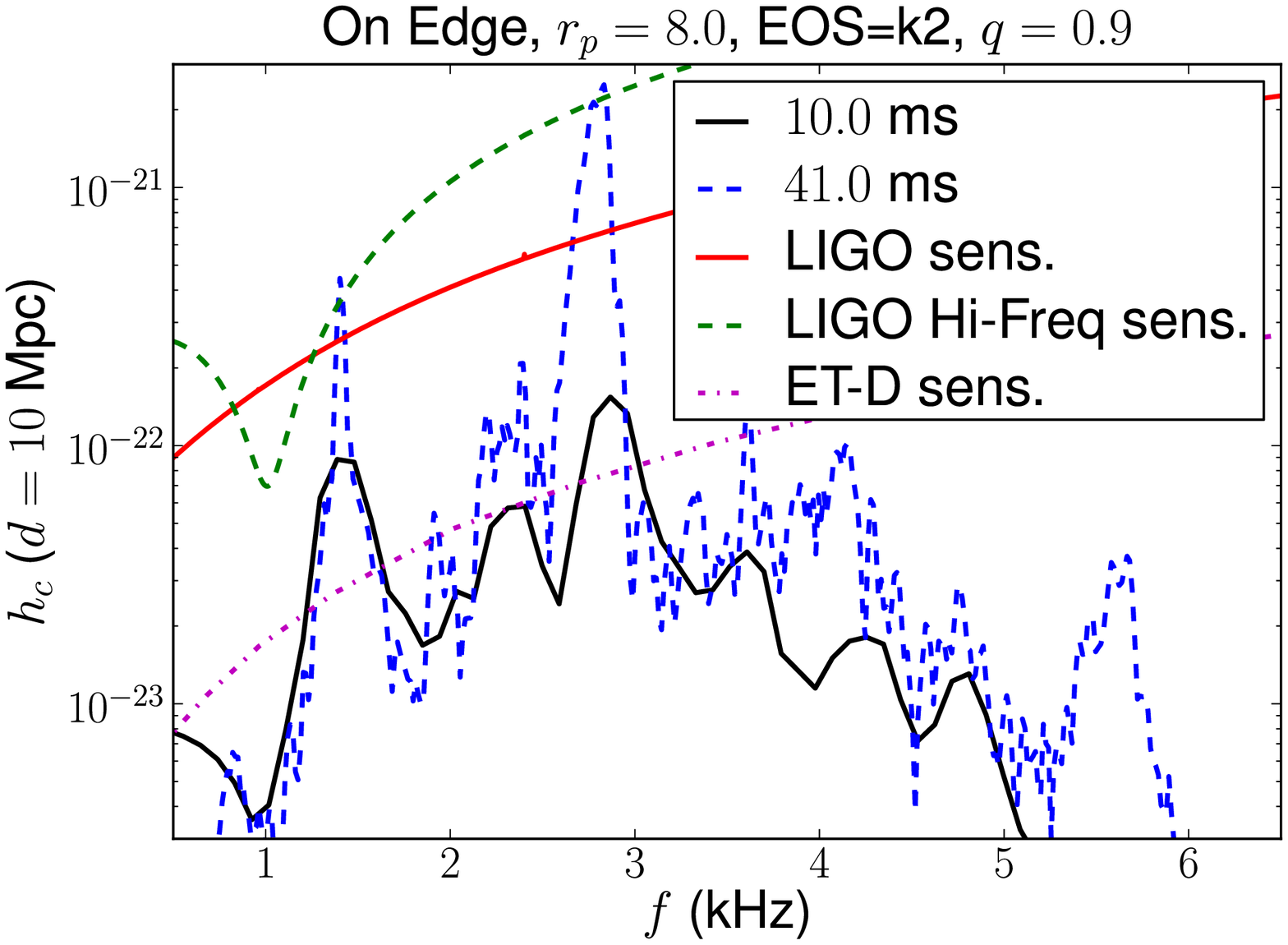}
\caption{ The characteristic gravitational wave strain as a function
  of frequency for an observer on edge comparing $\approx 10$ ms of
  the post-merger phase to the full time simulated ($\approx 41$ and
  65 ms, for the right and left panels, respectively) for a case with
  the 2H EOS and $q=0.9$ (left) and the $k_2$ EOS and $q=1$
  (right). In addition to the sensitivity curves for aLIGO and ET
  detectors, we also show a proposed ``High Frequency" configuration
  for aLIGO~\cite{ligo_noise}.  }\label{gwSpec_2}
\end{figure}

The hypermassive NSs can continue to emit GWs for much longer
times than the $10$ ms windows used for Fig.~\ref{gwSpec}, 
building more power in the frequency ranges associated with the
one-arm mode frequency and its integer multiples.  This is illustrated
in Fig.~\ref{gwSpec_2} where we compare the characteristic
gravitational wave strain for two cases, but for much longer
integration times.  From this it can be seen that after the violence
of the phase immediately following the merger, the GW power is concentrated into
very narrow frequency ranges, indicative of near monochromatic
emission at distinct frequencies, and the integrated power can grow significantly for a long-lived HMNS. In
Fig.~\ref{gwSpec_2} we have also included the sensitivity curve for a
proposed ``High Frequency" configuration for aLIGO that increases
sensitivity in a narrow range at around 1 kHz~\cite{ligo_noise}, that
happens to coincide with the $m=1$ mode frequency for the 2H EOS.
Hence, if future observations indicate a favorable binary NS merger
rate, such high frequency configurations may be worth considering to
learn more about the post-merger dynamics, and potentially reveal
information regarding the NS EOS.

We can estimate the approximate strength of the long-lived GW signals as follows. 
Assuming that: a) the source is observed on edge and b) the $m=1$ mode
has constant frequency and amplitude, the signal-to-noise ratio (SNR)
for the $m=1$ mode can be estimated via Eq. (81)
of~\cite{Jaranowski1998} (see also~\cite{2016arXiv160502369L}) and
approximating the 2,1 mode GW strain as $h_{21}\sim C_{21}/(2\pi
f_{m=1})^2$
%% \V{Please someone check the calculation for the SNR, I have redone it
%%   and I am getting a value much smaller than 0.7 now (namely 0.002).
%%   The main problem for the small number is the very small value for
%%   $C_{21}r$ we are getting compared to what Luis et
%%   al.~\cite{2016arXiv160502369L} find, who claim a value of $\sim
%%   10^{-4} (1/M_\odot)$ giving an SNR of 0.6 for similar parameters in
%%   the formula. Our maximum possible value is $C_{21}r \sim 10^{-3}
%%   (1/L)$ where $L$ is the unit length in our simulations, i.e.,
%%   $L=(2.7/0.01)*M_\odot = 270M_\odot$. Thus, in units of $1/M_\odot$
%%   our max value becomes $C_{21}r \sim 10^{-3}/270 \sim 4\times 10^{-6}
%%   (1/M_\odot)$, i.e. almost two orders of magnitude smaller than what
%%   Luis et al. claim. I don't see how they get such a high $C_{21}$
%%   amplitude (perhaps that very unequal mass mergers? I don't know..).
%%   However, it could be that I am doing something wrong here. So,
%%   please someone check the calculation. }
%
\labeq{}{ \hspace{-2.5cm}{\rm SNR}_{\rm aLIGO} \approx 2.8 \left(\frac{7\times 10^{-24}
    {\rm
      Hz}^{-1/2}}{\sqrt{S_n(f_{m=1})}}\right)\left(\frac{C_{21}r M}{10^{-4}}\right)\left(\frac{1.5
    {\rm kHz}}{f_{m=1}}\right)^{2}\left(\frac{T_{m=1}}{100 {\rm
      ms}}\right)^{1/2}\left(\frac{10{\rm Mpc}}{r}\right)
}
and
\labeq{}{ \hspace{-2.5cm}{\rm SNR}_{\rm ET} \approx 2.5 \left(\frac{8\times 10^{-25}
    {\rm
      Hz}^{-1/2}}{\sqrt{S_n(f_{m=1}})}\right)\left(\frac{C_{21}r M}{10^{-4}}\right)\left(\frac{1.5
    {\rm kHz}}{f_{m=1}}\right)^{2}\left(\frac{T_{m=1}}{100 {\rm
      ms}}\right)^{1/2}\left(\frac{100{\rm Mpc}}{r}\right) 
}
for the aLIGO zero-detuned high power configuration and the Einstein
Telescope (ET) ET-D configuration~\cite{ET}, respectively. Here,
$S_n(f_{m=1})$ is the detector noise spectral density at the frequency
of the one-arm mode and we adopt a mode lifetime of $T_{m=1}=100$ ms
(order of magnitude consistent with some of our simulations) and
distance to the source $r=10$ (100) Mpc for aLIGO (ET). Our SNR
predictions are less optimistic than those
in~\cite{2016arXiv160502369L}, but more optimistic than the
predictions in~\cite{2016arXiv160305726R}. However, we point out that
these papers focused on quasicircular binaries and adopted different
EOSs than we do. Note also that, as~\cite{2016arXiv160502369L} pointed
out, detection of a pre-merger NSNS signal will substantially lower
the SNR requirements to claim a detection of a post-merger $l=2$,
$m=1$ GW mode that is generated by a long-lived HMNS. Thus, the
one-arm mode could potentially be detectable by aLIGO at $10$ Mpc and
by ET at $100$ Mpc. However, if the one-arm mode survives intact for
about 1 s, then aLIGO could possibly detect such a signal to $\sim
30\rm Mpc$ and the ET to $\sim 300\rm Mpc$, which are similar
distances other studies have found for the $l=2$, $m=2$ GW modes
arising from post-merger oscillations (see
e.g.~\cite{Clark2016CQGra}). Note also that for softer equations of
state and unequal-mass mergers where $C_{21}rM\sim 10^{-3}$ the SNR
would be even larger, making the $m=1$ mode detectable for more
distant events. Different aLIGO configurations such as the ``High
Frequency'' one (see Fig.~\ref{gwSpec_2}) could also help through
lowering $S_n(f_{m=1})$ by a factor of a few. Such configurations
would be helpful in the future after observations have sufficiently
constrained the nuclear equation of state. However, all these
estimates depend on the assumptions that the source is observed
edge-on (although note that the $(2,1)$ mode amplitude is maximized
for an inclination $\theta=\pi/3$, at which the observed mode
amplitude is about 30\% larger than the edge-on ($\theta=\pi/2$)
case), and the signal frequency and amplitude are constant during the
lifetime of the mode. Therefore, the estimates are rather optimistic
because the mode frequency and amplitude are likely to change with
time, especially since magnetic fields and neutrino emission could
affect its properties. Nevertheless, our resolution study here and
in~\cite{PEPS2015,EPPS2016} demonstrate that at least when pure
hydrodynamical effects are considered, with higher resolution the
$m=1$ mode amplitude, once saturated, shows little to no signs of
decay during the span of the simulations. Longer, high-resolution
simulations that include detailed microphysics and magnetic fields are
necessary to accurately determine the long-term evolution of the
one-arm mode and detectability of the associated GWs.

These considerations indicate, as previous studies have also shown,
that it is worth searching for near quasistationary signals following
the merger of two neutron stars.  Measurement of the frequency of
these signals offers the potential to gain insights into the
properties of the structure of the HMNS, and matter above the nuclear
saturation density. Moreover, even the absence of a detection from a
relatively nearby source will allow one to constrain the nuclear EOS,
because the amplitude of the $m=1$ and $m=2$ modes vary significantly
with EOS when other system parameters are fixed. Of course,
degeneracies with other parameters would need to be considered, though
a more thorough sampling of parameter space is required in order for a
thorough investigation to be undertaken.

\subsection{Properties of unbound matter and electromagnetic counterparts}

\begin{figure}
\raggedleft
\includegraphics[trim =.0cm 0.cm 2.0cm  0.0cm,clip=true,width=0.49\textwidth]{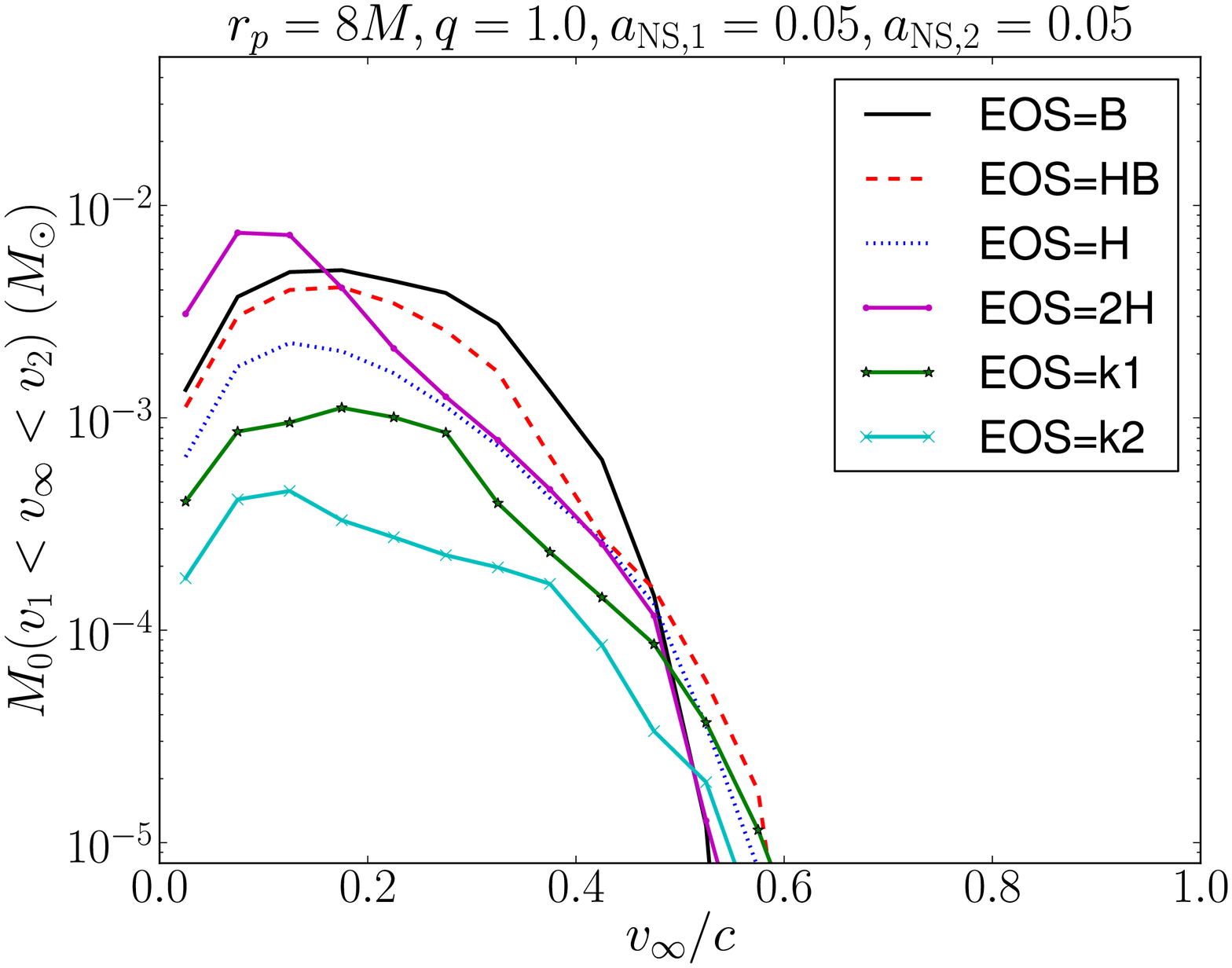}
\includegraphics[trim =.0cm 0.cm 2.0cm  0.cm,clip=true,width=0.49\textwidth]{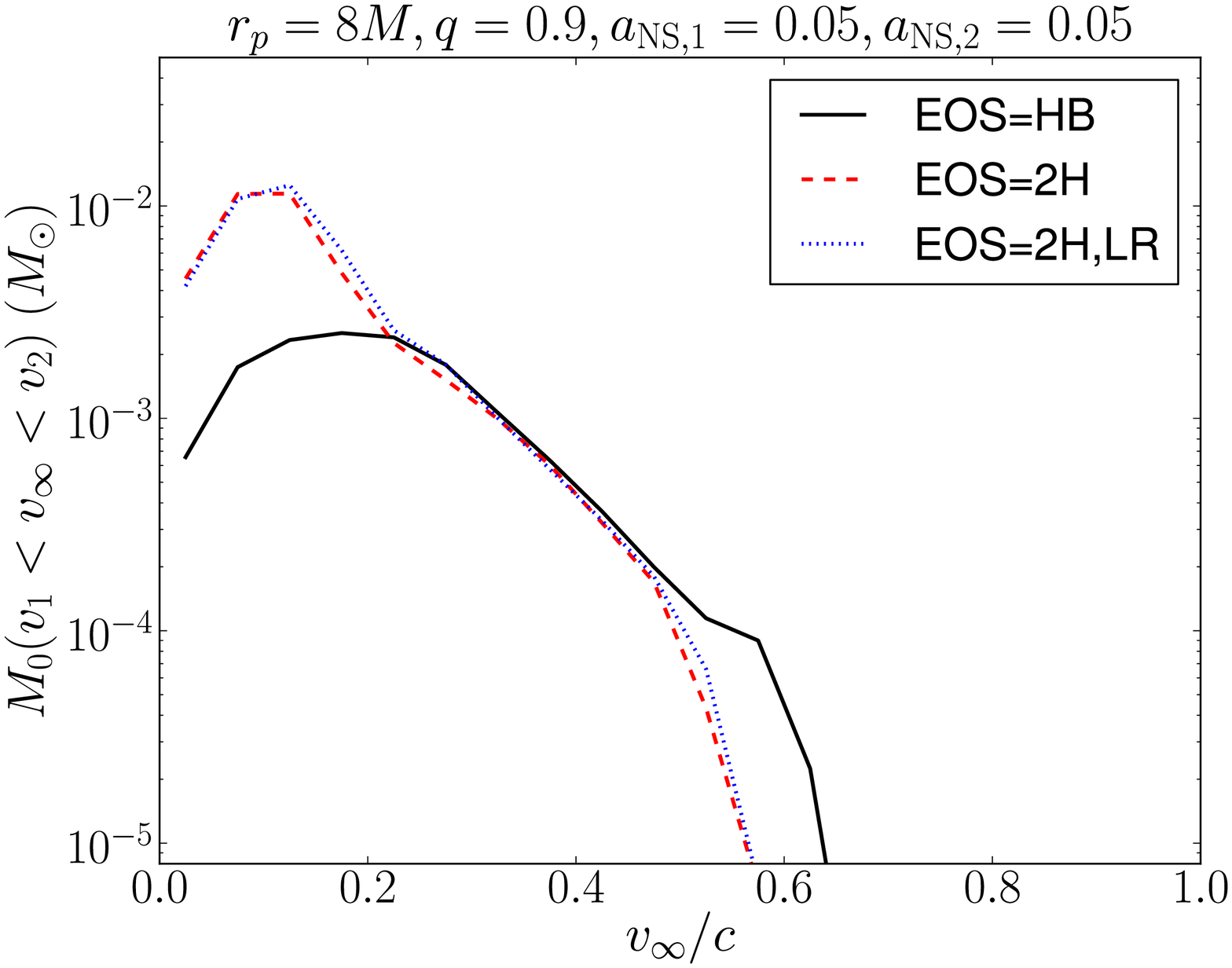}
\caption{Distribution of the asymptotic velocity of unbound rest-mass,
  binned in increments of 0.05c, and computed $\sim 9.3$ ms
  post-merger for several $r_p=8M$ cases. Left: $q=1.0$ for various
  EOSs. Right: $q=0.9$ for 2 EOSs, where we also include the results
  from a lower resolution (LR) run for the 2H EOS case.}\label{asym_vel}
\end{figure}

In Fig.~\ref{asym_vel} we show the distribution of the asymptotic
velocity of the ejected matter for several cases with periapse
$r_p=8M$, two different mass ratios and various EOSs. From this plot
one can see that, while there is substantial scatter, the stiffest
equation of state (2H) results in more mass ejected at lower
velocities than all other cases. In particular, amongst the $r_p = 8.0M$
runs, the 2H case with $q=0.9$ gives the highest yield of $\sim
0.015M_\odot$. However, the largest amount in ejected matter comes
from the $q=1.0$, $r_p = 9.5M$, 2H case with $\sim 0.026M_\odot$ of
ejected matter.

Based on these results we may estimate the properties of potential
kilonovae powered by fission of short-lived radioactive nuclei
produced through the r-process~\cite{Li:1998bw,2005astro.ph.10256K}.
Recent calculations of such processes suggest a rise time for
kilonovae lightcurves of~\cite{2013ApJ...775...18B}
\begin{equation}
t_{\rm peak}\approx0.3 \left(\frac{M_{\rm
0,u}}{10^{-2}M_{\odot}}\right)^{1/2}\left(\frac{v}{0.2c}\right)^{-1/2} \ \mbox{ d},
\label{tkilonovae}
\end{equation}
measured from the merger, and peak luminosities of
\begin{equation}L\approx 1.6\times 10^{41}\left(\frac{M_{\rm
0,u}}{10^{-2}M_{\odot}}\right)^{1/2}\left(\frac{v}{0.2c}\right)^{1/2}\mbox{
   erg $s^{-1}$}\label{Lkilonovae}.
\end{equation}
We list more details about the properties of the ejecta as well as of
potential electromagnetic counterparts in
Table~\ref{matter_table}. Note that a $L\sim10^{41} \rm erg\ s^{-1}$
kilonova at the edge of aLIGO's {\em quasi-circular} NSNS horizon
radius (at $\sim 200$ Mpc) would translate to an r-band magnitude of
23.5 mag~\cite{2013ApJ...775...18B}, one magnitude above the proposed
LSST survey sensitivity~\cite{2012arXiv1211.0310L}.  Hence factors of
a few difference in observed magnitude from such an intrinsic
luminosity may be discernible, and together with the GW signal could
allow one to constrain properties of the ejected matter, and
consequently parameters of the NSNS system that merged.  The highly
eccentric NSNS encounters studied here may only be detectable out to
roughly 1/10th the quasi-circular horizon distance~\cite{East:2012ww}
(larger impact parameter cases that result in multiple bursts could be
heard further away if templates or power stacking methods are used,
see ~\cite{East:2012xq,Tai:2014bfa}).  Of course, this means a
significantly lower expected event rate, but also that the
corresponding kilonova will have an apparent luminosity upwards of 100
times brighter than a 200 Mpc event.

\begin{table}[t]
\caption{\label{matter_table} Properties of the unbound material from
  various cases considered in this work. Listed are the periapse value
  of the initial orbit $r_p$, the equation of state, the dimensionless
  NS spins $a_{\rm NS,1}$ and $a_{\rm NS,2}$, the unbound rest mass
  $M_0$ in units of $M_\odot/100$, the rest-mass averaged asymptotic
  velocity $\langle v_{\infty}\rangle$, the total kinetic energy in
  units of $10^{50}$ erg, the anticipated kilonovae rise time and
  luminosity in units of d and $10^{41}$ erg/s, respectively, which we
  compute via Eqs.~\eqref{tkilonovae} and~\eqref{Lkilonovae}. We also
  include a lower resolution (2H,LR) run.  }  \centering
\begin{tabular}{cccccccccc}
\hline\hline
$r_p/M$ &
$q$ &
EOS &
$a_{\rm NS,1}$ &
$a_{\rm NS,2}$ &
$M_{0,\rm u}$ & 
$\langle v_{\infty}\rangle$ &
$E_{{\rm kin},50}$ &
$t_{\rm peak}$ &
$L_{41}$
\\
\hline
8.0   &   1.0   &   B   &  0.050   &   0.075   &   1.12   &   0.20   &   5.36   & 0.3 & 1.7 \\
8.0   &   1.0   &   H   &  0.050   &   0.075   &   0.44   &   0.19   &   1.95   & 0.2 & 1.0 \\
8.0   &   1.0   &   HB   &  0.050   &   0.075   &   0.84   &   0.19   &   3.65  & 0.3 & 1.4
\\
8.0   &   1.0   &   2H   &  0.050   &   0.075   &   1.08   &   0.14   &   2.69  & 0.4 & 1.4
\\
8.0   &   1.0   &   k1   &  0.050   &   0.075   &   0.24   &   0.20   &   1.18  & 0.1 & 0.8
\\
8.0   &   1.0   &   k2   &  0.050   &   0.075   &   0.09   &   0.20   &   0.49  & 0.1 & 0.5
\\
8.0   &   0.9   &   HB   &  0.050   &   0.050   &   0.56   &   0.21   &   3.01  & 0.2 & 1.2
\\
8.0   &   0.9   &   2H   &  0.050   &   0.050   &   1.52   &   0.13   &   3.51  & 0.5 & 1.6
\\
8.0   &   0.9   &   2H,LR   &  0.050   &   0.050   &   1.61   &   0.14   & 3.88 & 0.5 & 1.7
\\
9.5   &   1.0   &   2H   &  0.050   &   0.075   &   2.56   &   0.17   &   8.45  & 0.5 & 2.4
\\
\hline\hline 
\end{tabular}
\end{table}
%%%%%%%%%%%%%%%%%%%%%%%%%%%%%%%%%%%%%%%%%%%%%%%%%%%%%%%%%%%%%%%%%%%%%%%%%%%%

\section{Conclusions and Future Work}
\label{Conclusions}

In this work we considered eccentric mergers of binary neutron stars,
concentrating on how the equation of state and mass-ratio affect the
evolution of the resulting HMNS, and in particular the onset of the
one-arm instability.  Similarly
to~\cite{2016arXiv160305726R,2016arXiv160502369L}, we find that the
one-arm instability can occur in stars with stiff and soft EOSs, and
that the location of the narrowly peaked signal in frequency encodes
information about the NS EOS. We also find that binaries with
disparate mass-ratios can more readily seed larger one-arm modes in
the resulting HMNS star, in agreement with \cite{2016arXiv160502369L}.
Furthermore, we discovered that in addition to toroidal and
ellipsoidal HMNSs, double-core HMNSs can develop the $m=1$
instability. In this latter HMNS structure, strong $m=1$ density
perturbations can already be present in the double-core phase, due to
asymmetries in the two cores. Thus, correspondingly large $l=2$, $m=1$
gravitational-wave modes can arise even before the dominance of a
one-arm mode.  In addition, our results suggest that stiffer equations
of state produce HMNSs whose $m=1$ gravitational wave frequencies are
lower than HMNSs formed with softer equations of state. This suggests
that one should be able to find correlations between the frequency of
the $l=2$, $m=1$ GW mode and the EOS, in the same spirit as the
correlations that have been discovered for the $l=2$, $m=2$
mode~\cite{Stergioulas:2011gd,Takami2014,Takami2015,Bauswein2015a,Bauswein2015b,Bauswein2015c}.
Thus, if gravitational waves from the one-arm instability are
detected, they could in principle constrain the neutron star EOS. To
make this robust, quantitative predictions will require the study of a
larger suite of simulations with different equations of state, total
masses and mass ratios, all of which will be pursued in future work.

Through some simple estimates of the signal-to-noise ratio for aLIGO
and the future ET we concluded that, depending on the equation of
state, the one-arm mode could potentially be detectable by aLIGO at
$\sim 10$ Mpc and by ET $\sim 100$ Mpc. However, these estimates must
be checked against long, high-resolution simulations that account for
correct microphysics and magnetic fields, the impacts of which will be
the subjects of forthcoming works.

Finally, we have computed the properties of dynamically ejected matter
and find that at fixed periapse distance unequal-mass mergers eject
more matter and the associated kilonovae signatures are brighter than
in the case of equal-mass mergers (this is similar to the trends
with mass ratio seen in quasi-circular merger simulations). The upcoming LSST survey with its
exquisite sensitivity should be able to discern factors of a few
difference in luminosity of kilonovae and hence constrain the
properties of the ejected matter, and so potentially
constrain the parameters of the NSNS system that merged.

\ack

We thank Stuart Shapiro for access to the equilibrium rotating NS
code. This work was supported by NSF grant PHY-1607449, the Simons
Foundation, NASA grant NNX16AR67G (Fermi), and by Perimeter Institute
for Theoretical Physics. Research at Perimeter Institute is supported
by the Government of Canada through the Department of Innovation,
Science and Economic Development Canada and by the Province of Ontario
through the Ministry of Research, Innovation and Science.
Computational resources were provided by XSEDE/TACC under grant
TG-PHY100053, TG-MCA99S008 and the Orbital cluster at Princeton
University.

%\appendix

%\section{}

\section*{References}

\bibliographystyle{plain}
\bibliography{main}

\end{document}